\documentclass[11pt,a4paper]{article}
\usepackage{jcappub}

\title{Voids in Redshift Space}
\author[a]{Masatoshi Shoji}%
\author[b]{and Jounghun Lee}%

\affiliation[a]{Texas Cosmology Center and Department of Astronomy,
  The University of Texas at Austin,\\
  Austin, TX 78712, USA}
\affiliation[b]{Department of Physics and Astronomy, FPRD, Seoul National
  University, \\
  Seoul 151-747, Korea}
\emailAdd{mshoji@astro.as.utexas.edu}
\emailAdd{jounghun@astro.snu.ac.kr}

\abstract{
We study the ellipticity probability distribution function (PDF) of voids
in redshift space with galaxies as tracers of the shapes of voids.
We find that the redshift space distortion on the shape of voids
statistically increases the ellipticities of voids, and leaves
a prominent feature on the ellipticity PDF as a substantial reduction
in the probability of having voids with small ellipticity.
The location of this characteristic cutoff of the ellipticity PDF
is an explicit function of the logarithmic growth rate,
$f(z)\equiv\frac{d\ln D_+(z)}{d\ln a(z)}$,
and it can be used as a probe of cosmology once the radial density profile
of voids is better understood.
However, the biggest limiting factor for the use of ellipticity PDF as
a probe of cosmology lies in the Poisson noise from a small number of
galaxies to define the shape of a given void.
This Poisson noise creates a significant contamination
of the resulting ellipticity PDF so that the shape of the original PDF
is almost washed-out. Nevertheless, there is a way to overcome
the Poisson noise via the Alcock Paczynski test on the shape of
stacked voids. 
In redshift space, since the void is elongated toward the line of sight,
the stacked void has non-zero ellipticity, which can be a tell-tale of the
logarithmic growth rate.
Although some useful information of void ellipticity will be
lost by stacking, in this way, we can see the effect of redshift space
distortion as a source of anisotropy in the stacked
void ellipticity.
We think that the stacking analysis of the voids in redshift space
is potentially a powerful tool to probe the cosmology.
}
\begin{document}
\maketitle

\def\mnras{Mon. Not. R. Astron. Soc.}
\def\apj{Astrophys. Journal}
\def\apjl{Astrophys. J. Lett.}
\def\apjs{Astrophys. J. Suppl.}
\def\physrep{Phys. Rep.}
\def\aj{Astron. J.}
\def\aap{Astron. Astrophys.}
\def\pasj{Publ. Astron. Soc. Jap.}
\def\jcap{Journal of Cosmology and Astroparticle Physics}
\def\prd{Physical Review D}
\def\nat{Nature}
\section{Introduction}
The abundance of on-going and future galaxy surveys
is pointing to the deeper understanding of the nature of
accelerated expansion of the universe, as discovered
via observations of luminosity distances to Type Ia supernovae
\citep{riess/etal:1998,perlmutter/etal:1999}.
The main observables of galaxy surveys are characteristic
length scales encoded in the matter power spectrum, $P(k)$,
such as the comoving Hubble horizon, $H(z)$, and the
angular diameter distance, $D_A(z)$ \citep{shoji/jeong/komatsu:2009}.

Not only can we measure geometric distances, $D_{A}(z)$ and
$H(z)$, from a galaxy survey, but we also measure the growth rate of the
matter density fluctuation, $f(z)\equiv\frac{d\ln D_+(z)}{d\ln a(z)}$,
via redshift space distortions \citep{guzzo/etal:2008}.

The matter density fluctuation grows via gravitational instability
competing against the expansion of the universe.
One can obtain the growth rate by solving the following
differential equation \citep{wang/steinhardt:1998,linder/jenkins:2003,komatsu/etal:2009},
\begin{eqnarray}
&&\frac{d^2g}{d\ln a^2}
+\left[\frac52+\frac12\left(\Omega_k(a)-3w(a)\Omega_{DE}(a)\right)\right]\frac{dg}{d\ln a}
\nonumber\\
&&+\left[2\Omega_k(a)+\frac32\left(1-w(a)\right)\Omega_{DE}(a)\right]g(a)=0,
\label{eq:growth}
\end{eqnarray}
where
\begin{eqnarray}
g(a)&\equiv&\frac{D_+(a)}{a},\\
\Omega_k(a)&\equiv&\frac{\Omega_kH_0^2}{a^2H^2(a)},\\
\Omega_{DE}(a)&\equiv&\frac{\Omega_{DE}H_0^2}{a^{3[1+w_{\rm eff}(a)]}H^2(a)},\\
w_{\rm eff}(a)&\equiv&\frac1{\ln a}\int^{\ln a}_0d\ln a' w(a').
\end{eqnarray}
As we see in the eq. (\ref{eq:growth}), growth rate has an explicit dependence
on the dark energy equation of state.
Therefore, by measuring the growth rate, one can obtain the information
about the dark energy, thereby, the nature of accelerated expansion of the universe.

Predictions for the growth rate from different theories of gravity
and the nature of dark energy can be well parametrized
in the simple form of $f(z)=\Omega_m(z)^{\gamma}$.
For example, $\gamma=4/7$ for a $\Lambda$CDM model
\citep{peebles:1980,hamilton:2001},
where the cosmological constant plays the role of observed accelerated
expansion of the universe, and $\gamma=0.68$
for the DGP model of modifications of gravity \citep{lightman/schechter:1990}.
Thus, it is crucial to measure $f(z)$ from a given galaxy power spectrum
to observationally unveil the nature of this mysterious component
of the universe.
Although a galaxy power spectrum gives a way to measure
$f(z)$ at different redshifts, due to the vast parameter
space allowed, an accurate measurement of $f(z)$ from
a galaxy power spectrum alone is difficult \citep{simpson/peacock:2010}.
Especially, when we use the redshift space galaxy power spectrum
to the linear order,
it is inevitable to have the growth rate, $f(z)$, be degenerate
with a linear galaxy bias, $b_L$, such that
$P^s_g(k,\mu)=P_m(k,\mu)(1+\beta\mu^2)^2$, where $\beta(z)\equiv f(z)/b_L$
\citep{kaiser:1987}.
Thus, it is important to have an alternative method to measure
the growth rate in addition to a redshift space galaxy power spectrum.
Here, we propose the shape of voids in the redshift space
as a sensitive probe of linear growth rate, $f(z)$.

The shape of the voids in redshift space has been studied in literature
\citep{ryden:1995,ryden/melott:1996,maeda/sakai/triay:2011}.
According to their study, the void volume increases in the redshift space
as its boundary is stretched along the line of sight by redshift space distortion.
This elongation of the void volume in the redshift space is diametrically
opposite to what would happen to the distribution of galaxies in overdense
regions, due to a reversed sign of the equation of motion sourced
by local density field \citep{icke:1984}
(i.e., $\delta>0$ for over dense regions and $\delta<0$ for voids).
\citet{schmidt/ryden/melott:2001} studied the effect of
redshift space distortion on the void probability function (VPF).
The VPF is a measure of probability, $P_0(V)$, that a randomly placed
sphere of volume $V$ contains no galaxy within \citep{white:1979}.
For a given void in the real space, redshift space distortion
stretches the void along the line of sight, increasing the volume of voids;
thus, VPF, $P_0(V)$, increases toward a larger $V$.

Here, we study the {\it ellipticity} of voids in the redshift space
as a probe of the linear growth rate, $f(z)$. Specifically, we shall use
the ellipticity
probability distribution function of
\citet{park/lee:2007} (hereafter, PL07) and \citet{lee/park:2009}.
PL07 proposed a void ellipticity distribution function as a sensitive probe
of cosmology, noting that the shapes of voids are modulated by the
competition between the tidal distortion and the cosmic expansion.
Their model is based on the assumption that the underlying tidal field
can be well described by the Zel'dovich approximation.
They tested the model against the results from the Millennium
simulations \citep{springel/etal:2005}, finding a good agreement.

In practice, defining a void from a distribution of galaxies
is non-trivial, as the definition of a void differs from
one void finding algorithm to another
(\citet{colberg/etal:2008} and references there in).
Regardless of the ambiguity in the exact definition
and the boundary shape of a void,
the method of PL07 gives a robust measurement of ellipticities of voids
(i.e., defined as space around local density minima):
it uses the distribution of galaxies inside a void as a tracer
of the underlying shape of the tidal field.
Void statistics have been studied both in simulations and
galaxy redshift surveys
\citep{hoyle/vogeley:2002,patiri/etal:2006,foster/nelson:2009,pan/etal:2011}.
With the abundant data of the galaxy redshift survey available
from on-going and future galaxy surveys, properties of voids can be
better understood. Furthermore, voids found from these surveys
will provide a unique way to probe the history of the structure formation,
and henceforth, the nature of dark energy.


Throughout this paper we use the maximum likelihood cosmological
parameters of \citet{komatsu/etal:2009} (WMAP+BAO+SN of Table 1).

The goal of this paper is to provide an analytic insight into the
observed ellipticities of voids, and establish basis for their
application to cosmology.
In \S~\ref{sec:real_sp}, we briefly review notations and definitions used
in PL07 for a real space void ellipticity probability distribution function.
In \S~\ref{sec:bias}, we study the effect of linear galaxy bias on
the void ellipticity PDF.
In \S~\ref{sec:red_sp}, we study the effect of redshift space distortion
on the shape and ellipticity of observed voids.
In \S~\ref{sec:poisson_noise}, we study the effect of the Poisson noise on the
void ellipticity PDF, which arises as a consequence of tracing
a void shape by a limited number of field galaxies inside the void.
We also show that this Poisson noise can be the biggest limitation
on the use of PL07 method for voids with a small number of field galaxies
inside.
In \S~\ref{sec:n-body}, we compare the analytic prescriptions derived in the
previous sections against the void ellipticity PDF extracted from
the galaxy catalog of the Millennium simulation.
Finally, our conclusions are in \S~\ref{sec:diss_con}.

\section{Void in the real space}
\label{sec:real_sp}
In this section, we briefly review the method of PL07. 
First, voids are extracted from a given galaxy distribution using
a given void finding algorithm. We then find the ellipticities of voids,
and construct a void ellipticity distribution function.

For each detected void, we define void galaxies, which
reside inside a void, and calculate the inertia
tensor of equal weight using $N_{\rm vg}$ void galaxies,
\begin{eqnarray}
I_{ij}^r\equiv\sum_{\alpha=1}^{N_{\rm vg}}x_{i,\alpha}^rx_{j,\alpha}^r,
\label{eq:inertia_real}
\end{eqnarray}
where $x_i$ is the distance from the center of a void
(i.e., the mean, $\bar{x}_i$, has been removed).
Suppose that lengths of the semiaxis of the best-fit ellipsoid to a void
are $p_1$, $p_2$ and $p_3$, ordered as
$p_1\ge p_2\ge p_3\ge 0$. Then, one finds that they are proportional to the
square root of the eigenvalues of the inertia tensor, $I_1$, $I_2$ and $I_3$,
ordered as $I_1\ge I_2\ge I_3\ge 0$.
This motivates our defining the ellipticity of a void such that,
\begin{eqnarray}
\epsilon\equiv 1-\frac{p_3}{p_1}=1-\sqrt{\frac{I_3}{I_1}}
\nonumber\\
\eta\equiv 1-\frac{p_2}{p_1}=1-\sqrt{\frac{I_2}{I_1}}.
\label{eq:eps_eta}
\end{eqnarray}
For a fixed volume of an ellipsoid, $V=\frac{4\pi}3 p_1p_2p_3$,
we can exhaust all the possible shapes of the ellipsoid in terms of
$\epsilon$ and $\eta$.
\begin{figure*}[t]
\begin{center}
\rotatebox{0}{%
  \includegraphics[width=12cm]{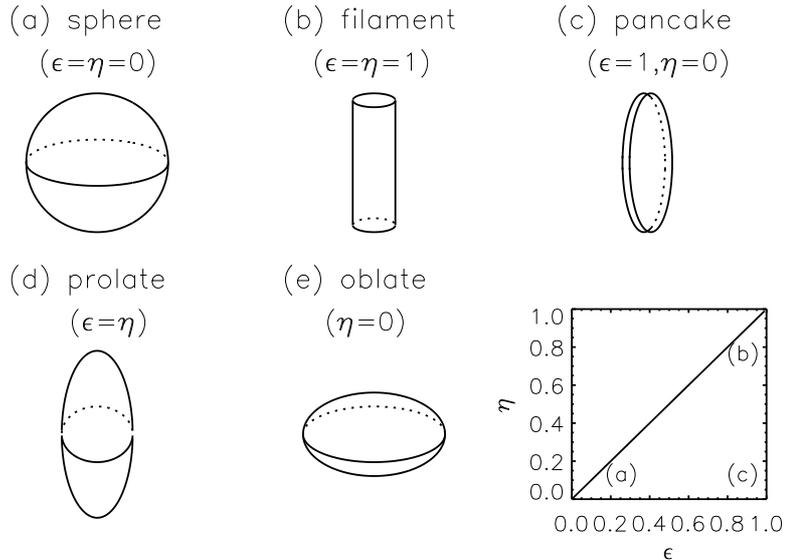}
}%
\caption{%
We show five special configurations of ellipsoids:
(a) sphere, (b) filament, (c) pancake, (d) prolate spheroid
and (e) oblate spheroid. In a plane of $\epsilon-\eta$,
the lower triangle shows the possible configurations of ellipsoid,
where each corner corresponds to a special configuration such as
a sphere (lower left), a filament (upper right) and a pancake (lower right).
Genuine triaxial ellipsoids are located in the middle of the triangle.
}%
\label{ellipsoids}
\end{center}
\end{figure*}
Figure \ref{ellipsoids} shows five special configurations of
ellipsoids on the $\epsilon$-$\eta$ plane.
Note that according to our definition (i.e., $p_1\ge p_2\ge p_3\ge 0$),
we have $0\le\eta\le\epsilon\le 1$.
\begin{enumerate}
  \renewcommand{\theenumi}{(\alph{enumi})}
  \renewcommand{\labelenumi}{\theenumi}
\item
  For $\epsilon=\eta=0$ (i.e., $p_1=p_2=p_3$), we have a spherical void.
\item
  For $\epsilon=\eta=1$ (i.e., $p_1\gg p_2=p_3=0$), we have a filamentary void.
\item
  For $\epsilon=1$ and $\eta=0$ (i.e., $p_1=p_2\gg p_3=0$), we have a pancake void.
\item
  For $\epsilon=\eta$ (i.e., $p_1>p_2=p_3$), we have a prolate spheroid.
\item
  For $\eta=0$ (i.e., $p_1=p_2>p_3$), we have an oblate spheroid.
\end{enumerate}
We define the triaxiality parameter following
\citet{franx/illingworth/dezeeuw:1991},
\begin{eqnarray}
T\equiv\frac{p_1^2-p_2^2}{p_1^2-p_3^2}=\frac{\eta(2-\eta)}{\epsilon(2-\epsilon)},
\end{eqnarray}
and further divide the triaxial ellipsoid into
two categories: prolate ellipsoids ($T>0.5$) and oblate ellipsoids ($T<0.5$).

\subsection{Real-Space Ellipticity PDF}
Here, we calculate the ellipticity probability distribution function
of PL07 based on the unconditional joint probability
density distribution of eigenvalues of a tidal tensor,
$\{\lambda_1,\lambda_2,\lambda_3\}$
\citep{doroshkevich:1970}.

An unconditional PDF is given by
\begin{eqnarray}
p(\lambda_1,\lambda_2,\lambda_3;\sigma_{R_L})
&=&\frac{3375}{8\sqrt{5}\pi\sigma_{R_L}^6}
\exp\left(-\frac{3K_1^2}{2\sigma_{R_L}^2}+\frac{15K_2}{2\sigma_{R_L}^2}\right)
\nonumber\\
&\times&(\lambda_1-\lambda_2)(\lambda_2-\lambda_3)(\lambda_1-\lambda_3),
\label{eq:doroshkevich}
\end{eqnarray}
where $K_1\equiv\lambda_1+\lambda_2+\lambda_3$,
$K_2\equiv\lambda_1\lambda_2+\lambda_2\lambda_3+\lambda_1\lambda_3$,
and $\sigma_{R_L}$ is the rms density fluctuation smoothed at the
Lagrangean void size, $R_L$, with a top-hat window function, $W(kR_L)$,
\begin{eqnarray}
\sigma_{R_L}^2\equiv\int^{\infty}_{-\infty}\Delta^2(k)W^2(kR_L)d\ln k.
\end{eqnarray}
Here, we calculate $\Delta^2(k)\equiv\frac{k^3P(k)}{2\pi^2}$
with a linear Boltzmann code, CAMB \citep{lewis/challinor/lasenby:2000},
with the best-fit cosmological parameters of WMAP+BAO+SN
\citep{komatsu/etal:2009}.

We derive a Lagrangean void size, $R_L$, from conservation of the number density
within the volume elements of Eulerian and Lagrangean space,
\begin{eqnarray}
n(\mathbf{x},z)d^3x=\bar{n}d^3q,
\end{eqnarray}
where $n(\mathbf{x},z)=\bar{n}(1+\delta(\mathbf{x},z))$ is
a real-space comoving number density of matter, $\bar{n}$ is
a mean comoving number density, and $\mathbf{x}$ and $\mathbf{q}$
are Eulerian and Lagrangean coordinates, respectively.
Therefore, we have
\begin{eqnarray}
R_L(\mathbf{q},z)=R_E(\mathbf{x},z)(1+\delta(\mathbf{x},z))^{1/3},
\end{eqnarray}
where $R_E$ is an Eulerian size of void.
PL07 derived the conditional PDF of $\{\lambda_1,\lambda_2\}$ for a given
density contrast, $\delta=\sum_{i=1}^3\lambda_i$ as follows,
\begin{eqnarray}
&&p(\lambda_1,\lambda_2|\delta,\sigma_{R_L})
=\frac{3375}{\sqrt{5\pi}\sigma_{R_L}^5}
\nonumber\\
&&\times\exp\left[
-\frac{5\delta^2}{2\sigma_{R_L}^2}
\left(1-\frac{3(\lambda_1+\lambda_2)}{\delta}
+\frac{3(\lambda_1^2+\lambda_1\lambda_2+\lambda_2^2)}{\delta^2}\right)
\right]
\nonumber\\
&&\times(2\lambda_1+\lambda_2-\delta)(\lambda_1-\lambda_2)
(\lambda_1+2\lambda_2-\delta).
\label{eq:pdf2d_lambda}
\end{eqnarray}
Here, as studied by \citet{lavaux/wandelt:2010}, signs of eigenvalues,
$\lambda_i$ tell us whether the void is spatially expanding or contracting
along the corresponding semiaxes.
In our notation, a given void spatially expands when $\lambda_i<0$,
and contracts when $\lambda_i>0$. Here, we only consider three dimensionally
expanding genuine voids, $0>\lambda_1\ge \lambda_2\ge \lambda_3$ and
$\delta=\sum^3_{i=1}\lambda_i<0$.

Now, under the premise of the strong correlation between
the void shape, $I_{ij}$, and the underlying tidal tensor,
$T_{ij}\equiv\frac{\partial^2\Phi(\mathbf{q})}{\partial q_i\partial q_j}$,
we have the following relations between the semimajor axis length, $p_i$,
and the eigenvalues of the tidal tensor, $\lambda_i$:
\begin{eqnarray}
\mu\equiv\frac{p_2}{p_1}=\left(\frac{1-\lambda_2}{1-\lambda_3}\right)^{1/2},\\
\nu\equiv\frac{p_3}{p_1}=\left(\frac{1-\lambda_1}{1-\lambda_3}\right)^{1/2},
\end{eqnarray}
where we assume $\lambda_1\ge \lambda_2\ge \lambda_3$.
Using these, one can define the PDF of $\epsilon$ and $\eta$ as
\begin{eqnarray}
p(\epsilon,\eta|\delta,\sigma_{R_L})
&=&p(\lambda_1(\mu,\nu),\lambda_2(\mu,\nu)|\delta,\sigma_{R_L})
\nonumber\\
&\times&\frac{4(\delta-3)^2\mu\nu}{(\mu^2+\nu^2+1)^3},
\label{eq:pdf2d}
\end{eqnarray}
where $\mu=1-\eta$ and $\nu=1-\epsilon$, and
\begin{eqnarray}
\lambda_1(\mu,\nu)=\frac{1+(\delta-2)\nu^2+\mu^2}{\mu^2+\nu^2+1}\\
\lambda_2(\mu,\nu)=\frac{1+(\delta-2)\mu^2+\nu^2}{\mu^2+\nu^2+1}.
\label{eq:defn_lambda}
\end{eqnarray}

From eq. (\ref{eq:doroshkevich}), we see that the probability of having
two of the three eigenvalues, $\lambda_i$, being equal is suppressed
due to the last three factors (i.e., $(\lambda_1-\lambda_2)(\lambda_2-\lambda_3)(\lambda_1-\lambda_3)$).
Therefore, we have small probabilities of having ellipsoids with
special configurations such as $\epsilon=\eta$ (prolate spheroids),
$\eta=0$ (oblate spheroids) and $\epsilon=0$ (sphere): these special
configurations correspond to $\lambda_1=\lambda_2$, $\lambda_2=\lambda_3$ and
$\lambda_1=\lambda_3$, respectively.
\begin{figure*}[t]
\begin{center}
\rotatebox{0}{%
  \includegraphics[width=14cm]{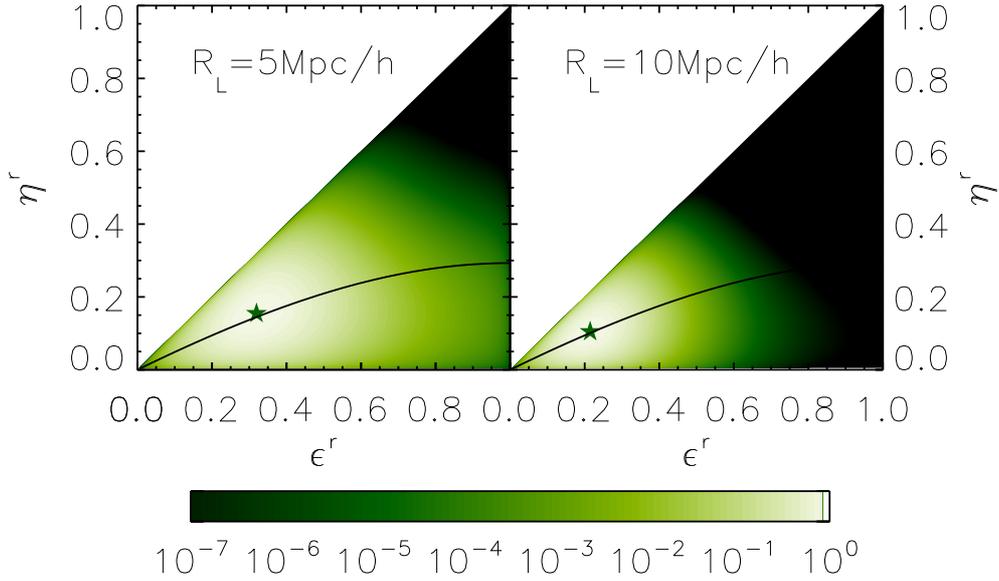}
}%
\caption{%
We show the ellipticity PDF for $R_L=5$ and $10h^{-1}{\rm Mpc}$ at $z=0$.
PDF is normalized so that the peak values are $1$. Solid lines show
$T=0.5$ and stars show the peaks of each PDF.
}%
\label{pdf2d_real}
\end{center}
\end{figure*}
In figure \ref{pdf2d_real}, we plot the ellipticity PDF of
eq. (\ref{eq:pdf2d}) for $R_L=5$ ({\it left}) and
$10~h^{-1}{\rm Mpc}$ ({\it right}), where $\delta_v=-0.9$.
PDF is normalized so that the peak value is unity (stars).
Note that for a fixed density contrast, $\delta$,
the exponential factor in eq. (\ref{eq:doroshkevich}) is maximized
when $\lambda_1=\lambda_2=\lambda_3$ allowing nearly spherical
voids to have a finite probability.

We also plot the line separating the prolate and oblate ellipsoids
(i.e., $T=0.5$) (solid line). We see that locations of the
ellipticity PDF peaks lie roughly on the line of $T=0.5$,
and PDF shifts toward smaller ellipticity both in
$\epsilon$ and $\eta$ for larger voids
(i.e., smaller $\sigma_{R_L}$ for a fixed set of cosmological parameters).

Also note that the ellipticity PDF in the real space
has almost no preference for prolate or oblate ellipsoids ($T\simeq 0.5$).
For $R_L=5h{\rm Mpc}^{-1}$, $\delta=-0.9$ and $z=0$, we have
mean ellipticities, $\bar{\epsilon}$ and $\bar{\eta}$ as follows,
\begin{eqnarray}
\bar{\epsilon}
&=&\int_0^1d\epsilon\int_{0}^{\epsilon/2} d\eta~ \epsilon~ p(\epsilon,\eta|\delta,\sigma_{R_L})
=0.37,\\
\bar{\eta}
&=&\int_0^1d\epsilon\int_{0}^{\epsilon/2} d\eta~ \eta~ p(\epsilon,\eta|\delta,\sigma_{R_L})
=0.16,\\
\bar{T}
&=&\int_0^1d\epsilon\int_{0}^{\epsilon/2} d\eta~ T(\epsilon,\eta)~ p(\epsilon,\eta|\delta,\sigma_{R_L})
=0.50,
\end{eqnarray}
and for $R_L=1h{\rm Mpc}^{-1}$, $\delta=-0.9$ and $z=0$,
we have $\bar{\epsilon}=0.62$, $\bar{\eta}=0.27$ and $\bar{T}=0.55$.

\begin{figure*}[t]
\begin{center}
\rotatebox{0}{%
  \includegraphics[width=10cm]{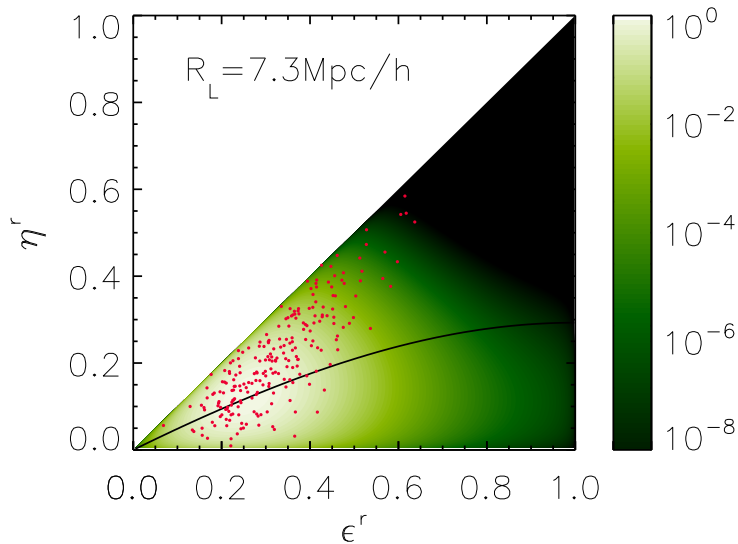}
}%
\caption{%
We show the 232 identified voids from the SDSS DR5
based on the catalog of \citet{foster/nelson:2009},
and the corresponding ellipticity PDF.
From the catalog, we find $\bar{\epsilon}=0.33$, $\bar{\eta}=0.21$,
$\bar{T}=0.68$, $\bar{\delta}_v=-0.97$, $\bar{z}=0.11$ and $R_L=7.3h^{-1}{\rm Mpc}$.
PDF is normalized so that the peak values are $1$, and the solid line shows $T=0.5$.
Note that the PDF is drawn only as a reference, and we do not correct for
bias and redshift space distortion.
}%
\label{pdf2d_FN}
\end{center}
\end{figure*}

However, previous observations and numerical studies indicate a preference
for prolate ellipsoids.
From voids detected in a $\Lambda$CDM N-body simulation,
\citet{platen/vandeweygaert/jones2008} claimed the axis ratio of best-fit
ellipsoids to be $1:0.7:0.5$ (i.e., $\epsilon=0.5$, $\eta=0.3$ and $T=0.68$),
and FN09 also found a preference for the prolate ellipsoids
from the voids detected in the SDSS DR5 with
$\epsilon=0.33$, $\eta=0.21$ and $\bar{T}=0.66$.
In figure \ref{pdf2d_FN}, we show the distribution of ellipticities
of 232 voids from the SDSS DR5 based on the catalog of
\citet{foster/nelson:2009}.
We also show the corresponding ellipticity PDF just as a reference.
The volume limited sample has a total of $52281$ galaxies
in a volume of $0.021h^{-3}{\rm Gpc^3}$
($\bar{n}_g\simeq 2.45\times 10^{-3}h{\rm Mpc^{-3}}$).
In the $\epsilon$-$\eta$ plane, we clearly see the dominance of prolate
voids, while analytic PDF has a slight to no preference of prolateness
in the shape of voids.
We think that the difference arises from the way we calculate ellipticities of voids.
Earlier works extracted void shapes by calculating the shape tensor from volume
elements inside detected voids. In this way, an extracted ellipticity is sensitive to the
boundary shape of the void as defined from a given void finder.
Since a boundary is defined from galaxies in an overdense region, $\delta\gtrsim0$,
the tidal field around the boundary is subject to a stronger distortion from
the external gravitational force.
An unconditional ellipticity PDF of \citet{doroshkevich:1970}, on the other hand,
describes a tidal field at and around the density minima.
Therefore, when we compare the analytic ellipticity PDF of PL07
against that of a simulation or galaxy survey,
it is important to know whether the extracted ellipticity truly
represents the shape of the tidal field around density minima.

\section{Linear Galaxy Bias in the Real-Space Void ellipticity PDF}
\label{sec:bias}
In the previous section, we have related the ellipticity PDF of voids to
the tidal field generated by the matter density distributions.
However, in reality, we obtain a void ellipticity PDF from a given galaxy distribution.
In order to deduce the underlying matter density field, $\delta_m$,
from galaxy distributions, we use the well known
linear galaxy bias, $b_L$, such that $\delta_v=b_L\delta_m$.
Here, $\delta_v$ represents the density contrast of a void calculated
from the number density of galaxies within the void, $n_{\rm vg}$,
and that of the cosmic mean, $\bar{n}_g$,
(i.e, average number density of galaxies in a survey volume),
\begin{eqnarray}
\delta_v\equiv \frac{n_{\rm vg}-\bar{n}_g}{\bar{n}_g}.
\end{eqnarray}
Since the unconditional PDF of \citet{doroshkevich:1970} gives
eigenvalues of tidal tensor, $T_{ij}$, generated by the matter density field,
we use $\delta_m=\delta_v/b_L$
in Eqs.(\ref{eq:pdf2d_lambda}) $\sim$ (\ref{eq:defn_lambda}),
where $b_L$ is a linear galaxy bias to be calibrated
from an observed ellipticity PDF.
The filtering scale, $R_L^m$, should also be calculated
following the definition of the linear bias,
\begin{eqnarray}
R_L^m&=&R_E^m(1+\delta_v/b_L)^{1/3}.
\end{eqnarray}
Here, an Eulerian size of void traced by the matter density field, $R_E^m$,
is not a direct observable in a galaxy survey
(i.e., the void size measured by the galaxy distribution,
$R_E^v$, is a direct observable).
\citet{patiri/etal:2006} showed that the radial density profiles of voids
defined from dark matter and halo distribution of N-body simulations
have similar shapes, and at large radii, haloes trace the dark matter.
Based on this result, we set $R_E^m\simeq R_E^v$ here, and have
\begin{eqnarray}
R_L^m&=&R_E^v(1+\delta_v/b_L)^{1/3}.
\label{eq:rl_bias}
\end{eqnarray}

\begin{figure*}[t]
\begin{center}
\rotatebox{0}{%
  \includegraphics[width=12cm]{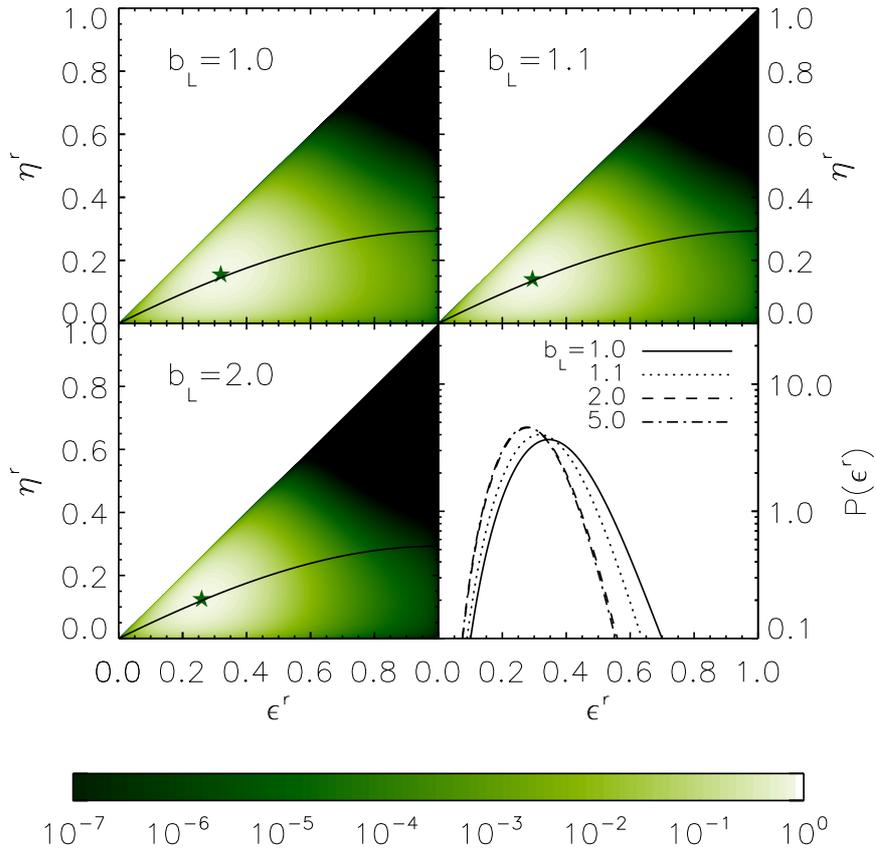}
}%
\caption{%
In this figure, we show the effect of linear galaxy bias on the
real space void ellipticity PDF both in 1-D ({\it bottom right}) and 2-D
({\it top left to bottom left}).
The 2-D PDFs are normalized so that the peak values are $1$.
In the 1-D plot, we have PDF for four different
biases, $b_L=1.0$ (solid), $1.1$(dotted), $2.0$ (dashed) and $5.0$ (dot-dashed).
In the 2-D plots, the solid lines show $T=0.5$, and the stars show the peaks
of each PDF.
Here, we have used $R_L=5h^{-1}{\rm Mpc}$, $\delta_v=-0.9$ and $z=0$.
}%
\label{pdf2d_bias}
\end{center}
\end{figure*}

Figure \ref{pdf2d_bias} shows the effect of linear galaxy bias on the
real space ellipticity PDF both in 1-D ({\it bottom right}) and in 2-D
({\it top left to bottom left}).
We normalize the 2-D PDFs to their peak values, and set the
parameters to $R_L=5h^{-1}{\rm Mpc}$, $\delta_v=-0.9$ and $z=0$.
In the 1-D plot, we have PDF for four different linear galaxy
biases, $b_L=1.0$ (solid), $1.1$(dotted), $2.0$ (dashed) and
$5.0$ (dot-dashed), and in the 2-D plots, we have
$b_L=1.0$ (top left), $1.1$ (top right) and $2.0$ (bottom left).

In the 2-D plots, the solid lines show $T=0.5$, and the stars show the peaks
of each PDF. Regardless of the values of bias, peaks of PDF lie on
the solid lines ($T=0.5$), and the ratio of prolate to oblate spheroids is
$\sim 0.5$ as in the case of unbiased (i.e., $b_L=1$) ellipticity PDF.

We clearly see an effect of bias on the void PDF as we increase
the galaxy bias $b_L$ from its unbiased value, $b_L=1$, to biased value, $b_L> 1$.
As we increase the bias from $b_L=1$, the ellipticity PDF peaks
at smaller ellipticities more sharply. For $b_L>2.0$, the shape of PDF converges
quickly, and becomes insensitive to the further change in $b_L$.
As a larger bias reduces the matter density contrast, $\delta_m$,
for a fixed value of $\delta_v$, a smaller
matter density contrast yields a less pronounced tidal field around
the density minima, yielding smaller ellipticity.

In Eqs.(\ref{eq:pdf2d_lambda}) $\sim$ (\ref{eq:defn_lambda}),
we see the effect of bias through $\sigma_{R_L}$.
Since the Lagrangean size of void, $R_L^m$,
(as defined in eq. (\ref{eq:rl_bias})) sets the smoothing scale
of the density field for calculating $\sigma_{R_L}$, the larger the
$R_L^m$ is, the smaller the $\sigma_{R_L}$ becomes.
Again, a smaller rms fluctuation
of the matter density field leads to a weaker tidal field, and hence, the
smaller ellipticity of the void.
We see that, although the linear bias, $b_L$, and $\sigma_8$,
which is directly proportional to the $\sigma_{R_L}$,
are nearly degenerate in the parameter space of the
biased ellipticity PDF, their correlation is directly opposite
in the case of the galaxy power spectrum, $P_g(k)\propto b_L^2\sigma_8^2$.
This feature of the biased void ellipticity PDF allows us to
better constrain both the linear bias, $b_L$, and the rms density fluctuations,
$\sigma_8$, when combined with the galaxy power spectrum.

\section{Void in the redshift space}
\label{sec:red_sp}
In order to compare the analytic and observed ellipticity PDFs,
we need to understand the effect of the redshift space distortion
on the ellipticity of a given void.
Unlike galaxies and galaxy clusters,
which reside in density maxima, voids reside in density minima,
where the diverging velocity field pushes the boundary and the field
galaxies inside the voids outward, expanding their volume in all
directions \citep{icke:1984}.

A real-space inertia tensor is defined in
eq. (\ref{eq:inertia_real}),
and that of the redshift-space is similarly defined as
\begin{eqnarray}
I^s_{ij}\equiv\sum^{N_{\rm vg}}_{\alpha=1}
x_{\alpha,i}^sx_{\alpha,j}^s,
\label{eq:inertia_red}
\end{eqnarray}
where $N_{\rm vg}$ is the number of void galaxies, and $\mathbf{x}^s_{\alpha}$
is the redshift-space position of an $\alpha$-th void galaxy.
An inertia tensor is symmetric by definition (i.e., $I_{ij}=I_{ji}$), and
traces a shape of a void from an underlying void galaxy distribution.
\citet{park/lee:2007} identified voids from the Millennium simulation using the void
finding algorithm of \citet{hoyle/vogeley:2002},
and extracted the ellipticity PDF of voids using the inertia tensor
using void galaxies as tracers.
They found that the distribution of ellipticities of voids
from the Millennium simulation follows that of linear theory
(Zel'dovich Approximation) prediction in real space.

Here, we study the effect of the redshift space distortion on the
ellipticity PDF.
First, we calculate the redshift space inertia tensor as follows.

A mapping between the real-space and the redshift space is given by
\begin{eqnarray}
x_1^s&=&x_1^r
\label{eq:x1}\\
x_2^s&=&x_2^r
\label{eq:x2}\\
x_3^s&=&x_3^r-f\mathbf{u}\cdot\hat{\mathbf{x}}_3^r,
\label{eq:x3}
\end{eqnarray}
where $f\equiv\frac{d\ln\delta(a)}{d\ln a}$ is the growth rate,
and the unit vector, $\hat{\mathbf{x}}^r_3$, is along the
line of sight.
Here, $\mathbf{u}\equiv\frac{-\mathbf{v}}{\mathcal{H}f}$, where
$\mathbf{v}$ is the peculiar velocity of a void galaxy,
and $\mathcal{H}(a)=aH(a)$ is a comoving Hubble rate.

We define $\kappa(\mathbf{x}^r)$ to quantify the strength of
the redshift space distortion as
\begin{eqnarray}
\kappa(\mathbf{x}^r)\equiv
-\frac{f\hat{\mathbf{x}_3^r}\cdot\mathbf{u}(\mathbf{x}^r)}{x_3^r}.
\end{eqnarray}
With $\kappa(\mathbf{x}^r)$, we re-write eq. (\ref{eq:x3})
as
\begin{eqnarray}
x_3^s&=&x_3^r(1+\kappa(\mathbf{x}^r)),
\label{eq:x3b}
\end{eqnarray}
and we relate the redshift-space inertia tensor to the real-space
inertia tensor as follows,
\begin{eqnarray}
I^s_{11}&=&I^r_{11}
\label{eq:I11s}\\
I^s_{12}&=&I^r_{12}\\
I^s_{22}&=&I^r_{22}\\
I^s_{13}&=&I^r_{13}+\sum^{N_{\rm vg}}_{\alpha=1}
x_{\alpha,1}^rx_{\alpha,3}^r\kappa_{\alpha}\\
I^s_{23}&=&I^r_{23}+\sum^{N_{\rm vg}}_{\alpha=1}
x_{\alpha,2}^rx_{\alpha,3}^r\kappa_{\alpha}\\
I^s_{33}&=&I^r_{33}+2\sum^{N_{\rm vg}}_{\alpha=1}
x_{\alpha,3}^rx_{\alpha,3}^r\kappa_{\alpha}+\sum^{N_{\rm vg}}_{\alpha=1}
x_{\alpha,3}^rx_{\alpha,3}^r\kappa_{\alpha}^2,
\label{eq:I33s}
\end{eqnarray}
where
\begin{eqnarray}
\kappa_{\alpha}=-\frac{f\mu_{\alpha}u_{\alpha}}{x_{\alpha,3}^r},
\label{eq:kappa_i}
\end{eqnarray}
and $\mu$ is the angle between
the peculiar velocity, $\mathbf{u}$, and the line of sight
(i.e., $\mu\equiv\hat{\mathbf{u}}\cdot \hat{x}_3^r$).

Once we have the redshift space inertia tensor either numerically,
from eq. (\ref{eq:inertia_red}), or analytically in terms of $\kappa$,
from eqs. (\ref{eq:I11s}) $\sim$ (\ref{eq:I33s}), we can calculate the
ellipticity of a given void, $\epsilon$ and $\eta$, from eigenvalues
of the redshift space inertia tensor, ($I_1^s\ge I_2^s\ge I_3^s$)
as in eq. (\ref{eq:eps_eta}).

Since $\kappa$ is a function of the peculiar velocity field,
we can use linear theory to calculate $\kappa$ for an arbitrary
density field, $\delta(\mathbf{x}^r)$, as,
\begin{eqnarray}
\kappa(\mathbf{x}^r)&\equiv&
-\frac{f\hat{\mathbf{x}_3^r}\cdot\mathbf{u}(\mathbf{x}^r)}{x_3^r}
=-\frac{f\hat{\mathbf{x}_3^r}\cdot\mathbf{\nabla}\Phi(\mathbf{x}^r)}{4\pi Ga^2\bar{\rho}x_3^r}
\nonumber\\
&=&-\frac{f}{4\pi Ga^2\bar{\rho}x_3^r}
\int\frac{d^3k}{(2\pi)^3}i\mathbf{k}\cdot\hat{\mathbf{x}}_3^r
\tilde{\Phi}(\mathbf{k})e^{i\mathbf{k}\cdot\mathbf{x}^r}
\nonumber\\
&=&\frac{f}{x_3^r}\int\frac{d^3k}{(2\pi)^3}i\mathbf{k}\cdot\hat{\mathbf{x}}_3^r
\frac{\tilde{\delta}(\mathbf{k})}{k^2}e^{i\mathbf{k}\cdot\mathbf{x}^r}
\nonumber\\
&=&\frac{f}{x_3^r}\int_0^{\infty}\frac{dk}{(2\pi)^3}ik\int_0^{2\pi}d\phi_k\int_{-1}^1
d\mu_k \mu_k\tilde{\delta}(\mathbf{k})e^{i\mathbf{k}\cdot\mathbf{x}^r},
\nonumber\\
\label{eq:kappa}
\end{eqnarray}
where $\mu_k\equiv\cos\theta_k$.
Here, we used the linear relation between a peculiar velocity and
a gravitational field derived from Zel'dovich Approximation
\citep{hui/bertschinger:1996},
\begin{eqnarray}
\mathbf{v}=-\frac{\mathcal{H}f}{4\pi Ga^2\bar{\rho}}\mathbf{\nabla}\Phi,
\end{eqnarray}
and a Poisson equation,
\begin{eqnarray}
\nabla^2\Phi=4\pi Ga^2\bar{\rho}\delta.
\end{eqnarray}
We explicitly write a cosine angle between $\mathbf{k}$ and $\mathbf{x}$,
in a spherical coordinate as
\begin{eqnarray}
\cos\gamma\equiv \cos\theta_k\cos\theta_x+\cos(\phi_k-\phi_x)\sin\theta_k\sin\theta_x,
\end{eqnarray}
where $\mathbf{k}\cdot\mathbf{x}=kx\cos\gamma$.
Here, in order to calculate the effect of peculiar velocity on the void shape,
we need to know the density profile of a void, $\delta(\mathbf{x})$ in the real space.
\citet{sheth/vandeweygaert:2004} studied an evolution of a radial
density profile of an isolated void, and found that the density
profile evolves into an increasingly similar shape to a top-hat function.
With N-body simulations both in
sCDM and $\Lambda$CDM, \citet{colberg/etal:2005} found that the
radial density profile of a void is universal, having
$\rho(r<r_{\rm eff})/\rho(r_{\rm eff})\propto \exp[(r/r_{\rm eff})^{\alpha}]$,
where $r_{\rm eff}$ is the effective radius of the void and $\alpha\sim 2$.
These analytic radial density profiles are also seen in the voids
identified from the galaxy distribution of the Sloan Digital Sky Survey
as a sharp rise of density contrast near the boundary of the void
\citep{patiri/etal:2006,pan/etal:2011}.
In this section, we approximate the radial density profile of the void, $\delta(r)$,
to be a top-hat function for simplicity,
\begin{eqnarray}
\delta(\mathbf{x}^r)=\left\{
\begin{array}{c}
\bar{\delta}~~~{\rm within~the~void}
\\
0~~~{\rm outside~the~void}
\end{array},\right.
\end{eqnarray}
where $\bar{\delta}$ is the average density contrast within the void.
We will discuss further about more general density profiles at the end of this section.

Below, we will show analytic solutions for the redshift space ellipticity for some of
the limiting cases.

\subsection{Spherical Void}
First, we consider a sphere with zero real space ellipticity,
$\epsilon^r=0$ and $\eta^r=0$, with three eigenvalues of the
inertia tensor of $I_1^r=I_2^r=I_3^r$.
For a spherical top hat density field, we have
\begin{eqnarray}
\tilde{\delta}(\mathbf{k})=4\pi R^3\bar{\delta}\frac{j_1(kR)}{kR}.
\end{eqnarray}
Therefore, eq. (\ref{eq:kappa}) becomes
\begin{eqnarray}
\kappa(\mathbf{x}^r)&=&\frac{4\pi R^3f\bar{\delta}}{x_3^r}
\int_0^{\infty}\frac{dk}{(2\pi)^3}ik\frac{j_1(kR)}{kR}
\nonumber\\
&\times&\int_0^{2\pi}d\phi_k\int_{-1}^1d\mu_k \mu_k e^{i\mathbf{k}\cdot\mathbf{x}^r}.
\end{eqnarray}
We first integrate the angular portion of the integration,
$\int d\Omega_k \mu_k e^{i\mathbf{k}\cdot\mathbf{x}^r}$,
by expanding the exponent into Legendre polynomials,
\begin{eqnarray}
e^{i\mathbf{k}\cdot\mathbf{x}^r}=\sum_l^{\infty}(-i)^l(2l+1)j_l(-kx)P_l(\cos\gamma),
\end{eqnarray}
and using the Spherical harmonics addition theorem (a.k.a., Legendre addition theorem). We find
\begin{eqnarray}
&&\int d\Omega_k \mu_k e^{i\mathbf{k}\cdot\mathbf{x}^r}
=4\pi ij_1(kx)\cos\theta_x.
\label{eq:solid_angle_sphere}
\end{eqnarray}
Finally, we obtain $\kappa$ for a spherical void as
\begin{eqnarray}
\kappa(\mathbf{x}^r)&=&\frac{4\pi R^3f\bar{\delta}}{x_3^r}
\int_0^{\infty}\frac{dk}{(2\pi)^3}ik\frac{j_1(kR)}{kR}
[4\pi ij_1(kx)\cos\theta_x]
\nonumber\\
&=&-\frac{2R^2f\bar{\delta}\cos\theta_x}{\pi x_3^r}
\int_0^{\infty}dk j_1(kR)j_1(kx)
\nonumber\\
&=&-\frac{2R^2f\bar{\delta}\cos\theta_x}{\pi x_3^r}\left(\frac{\pi x}{6R^2}\right)
=-\frac{f\bar{\delta}}{3}.
\label{eq:sphere_kappa}
\end{eqnarray}
As we see from eq. (\ref{eq:sphere_kappa}), $\kappa$ is independent of
positions in the void, and therefore the three eigenvalues of
the redshift space inertia tensor for a redshift space void,
$I_1^s\ge I_2^s\ge I_3^s$, are
\begin{eqnarray}
I_1^s&=&I_1^r (1+\kappa)^2\\
I_2^s&=&I_2^r\\
I_3^s&=&I_3^r,
\end{eqnarray}
and the ellipticity, $\epsilon$ and $\eta$ are
\begin{eqnarray}
\epsilon^s=\eta^s=1-\frac1{1+\kappa}=1-\frac1{1-f\bar{\delta}/3},
\label{eq:eps_eta_sphere}
\end{eqnarray}
following eq. (\ref{eq:eps_eta}).
Now, the linear growth rate, $f\equiv\frac{d\ln D(a)}{d\ln a}$,
for the $\Lambda$CDM model is approximately given as \citep{peebles:1980,hamilton:2001},
\begin{eqnarray}
f(z)=\Omega(z)^{4/7},
\end{eqnarray}
where
\begin{eqnarray}
\Omega(z)=\frac{\Omega_m(1+z)^3}{\Omega_m(1+z)^3+\Omega_{\Lambda}}.
\end{eqnarray}
For $\Omega_m=0.25$ and $\Omega_{\Lambda}=0.75$, $f\simeq 0.5$ at $z=0$,
and thus, for a spherical void with an average density contrast of
$\bar{\delta}=-0.9$, we have $\epsilon^s=\eta^s\simeq 0.13$.
Since there is no special orientation for a spherical void,
any void with sufficiently small real space ellipticity,
$\epsilon^r\sim\eta^r\sim 0$, becomes a prolate ellipsoid
(i.e., $\epsilon^s\sim\eta^s\ne 0$) in the redshift space with
its ellipticity given by eq. (\ref{eq:eps_eta_sphere}).

\subsection{Spheroidal Void}
When two of the three lengths of the semiaxis are equal,
we have a spheroidal void. We set the direction of the longest
and shortest semiaxis to be along the line of sight for prolate
and oblate spheroids respectively.
As in the spherical void, we assume a spheroidal top hat density
contrast and solve eq. (\ref{eq:kappa}) to have $\kappa$ to the
first order in $\epsilon^r$,
\begin{eqnarray}
\kappa(\mathbf{x}^r)\simeq -\frac{f\bar{\delta}}3(1\mp2\epsilon^r,)
\end{eqnarray}
where ``$-$'' and ``$+$'' signs are for prolate and oblate spheroids
respectively.
Here, we set the longest axis along the line of sight for
prolate spheroids, and set the shortest axis along the line of sight
for oblate spheroids.
We compared the analytic prediction against numerical calculations,
and found that
\begin{eqnarray}
\kappa(\mathbf{x}^r)\simeq -\frac{f\bar{\delta}}3(1\mp\epsilon^r,)
\label{eq:kappa_pro_ob_fit}
\end{eqnarray}
fits the result better.
For each case, we have redshift space ellipticity as a function
of real space ellipticity and $\kappa$ as follows,
\begin{eqnarray}
\epsilon^s=1-\frac{1-\epsilon^r}{1+\kappa},\\
\label{eq:eps_s_pro}
\epsilon^s=1-(1-\epsilon^r)(1+\kappa).
\label{eq:eps_s_ob}
\end{eqnarray}
In contrast to the spherical void, whose redshift space ellipticity
is always larger than that of real space, ellipticity of
a spheroid either increases or decreases in the redshift space.

\begin{figure*}[t]
\begin{center}
\rotatebox{0}{%
  \includegraphics[width=14cm]{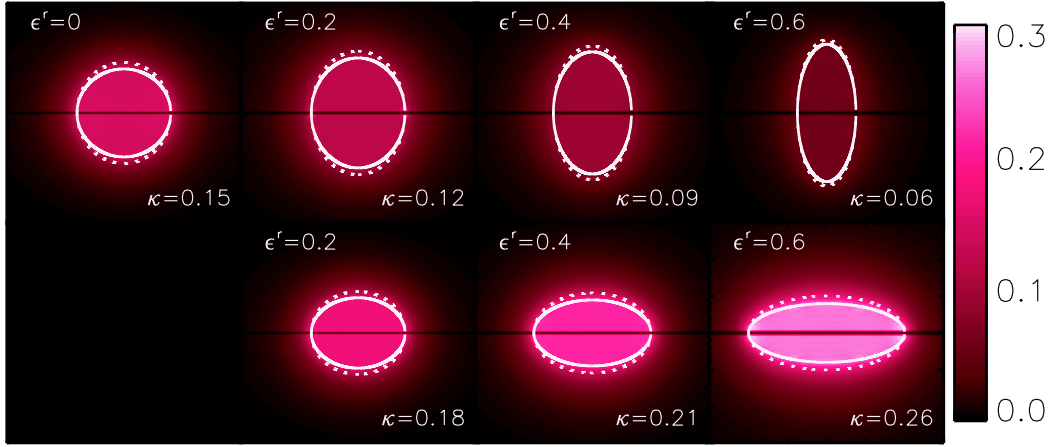}
}%
\caption{%
We show the spatial distribution of
the value of $\kappa(\mathbf{x}^r)$ in x-z plane for prolate spheroids
with its longest axis along the line of sight ({\it top}),
and oblate ellipsoids with its shortest axis along the line of sight
({\it bottom}) for different real space ellipticities, $\epsilon^r$.
We also plot real space (solid lines) and redshift space (dotted lines)
shapes of voids. Here, we use $f=0.5$ and $\delta=-0.9$.
}%
\label{kappa_contour}
\end{center}
\end{figure*}

In figure \ref{kappa_contour}, we show the result of numerical calculations
for the spatial distribution of
the value of $\kappa(\mathbf{x}^r)$ in the x-z plane for prolate spheroids
with its longest axis along the line of sight ({\it top}),
and oblate ellipsoids with its shortest axis along the line of sight
({\it bottom}) for different real space ellipticities, $\epsilon^r$.
In the figure, the center of the spheroids corresponds to the origin of the
coordinate, and the x-z plane cuts the middle of the spheroid.
We also plot the real space (solid lines) and redshift space (dotted lines)
shapes of voids. Here, we use $f=0.5$ and $\delta=-0.9$.

We see that the deformation of the void shape, $\kappa\propto u_3$,
is the largest when the shortest semiaxis is along the line of sight
due to the largest line-of-sight peculiar velocity.
This result is in agreement with the earlier work of \citet{icke:1984}.
When the longest semiaxis is along the line of sight (i.e., {\it top} figures),
ellipticity only {\it increases} in the redshift space,
while ellipticity only {\it decreases} when the shortest semiaxis is along
the line of sight (i.e., {\it bottom} figures).
For the spherical void, $\epsilon^r=0$, we have $\epsilon^s=0.13$
as expected from eq. (\ref{eq:eps_eta_sphere}).
For prolate and oblate spheroids with $\epsilon^r=0.2$, $0.4$ and $0.6$,
we have $\kappa=0.12$, $0.09$ and $0.06$, and
$\kappa=0.18$, $0.21$ and $0.26$ from the numerical calculations respectively.
This is in a good agreement with the fitting function of
eq. (\ref{eq:kappa_pro_ob_fit}):
for prolate and oblate spheroids with $\epsilon^r=0.2$, $0.4$ and $0.6$,
eq. (\ref{eq:kappa_pro_ob_fit}) yields $\kappa=0.12$, $0.09$ and $0.06$, and
$\kappa=0.18$, $0.21$ and $0.24$, respectively.

Using Eqs.(\ref{eq:eps_s_pro}) and (\ref{eq:eps_s_ob}),
for prolate and oblate spheroids with $\epsilon^r=0.2$, $0.4$ and $0.6$,
we have redshift space ellipticities of $\epsilon^s=0.29$, $0.45$ and $0.62$,
and $\epsilon^s=0.06$, $0.27$ and $0.50$ respectively.
In general, the fractional change in the ellipticity,
$(\epsilon^s-\epsilon^r)/\epsilon^r$, becomes smaller for a larger $\epsilon^r$.

Although we have constant $\kappa$ here (i.e., independent of the position
in the void), this is not true for arbitrarily oriented voids.
In the case of tilted voids, $\kappa(\mathbf{x}^r)$ is a function of
the real space coordinate, $\mathbf{x}^r$, and has both positive and
negative values by definition, $\kappa(\mathbf{x}^r)\propto u_3/x_3$.
\subsection{Redshift-Space Ellipticity Distribution Function}

For a given ellipticity in real space, the corresponding redshift-space
ellipticity can either increase or decrease, depending on the real
space shape and orientation of the void with respect to the line of sight,
except for a spherical void, whose ellipticity can only increase.
Here, we calculate changes in ellipticities
for a given set of real space ellipticities, $\epsilon^r$ and $\eta^r$,
\begin{eqnarray}
\Delta\epsilon(\epsilon^r,\eta^r)\equiv\epsilon^s(\epsilon^r,\eta^r)-\epsilon^r\\
\Delta\eta(\epsilon^r,\eta^r)\equiv\eta^s(\epsilon^r,\eta^r)-\eta^r,
\end{eqnarray}
averaged over possible orientations.
We numerically solve eq. (\ref{eq:kappa}) for
$\kappa(\mathbf{x}^r)$, with a top-hat density contrast,
$\delta(\mathbf{x}^r)$, and calculate the redshift-space
ellipticity, $\epsilon^s$ and $\eta^s$, from the redshift-space
inertial tensor, $I^s_{ij}$.
\begin{figure*}[t]
\begin{center}
\rotatebox{0}{%
  \includegraphics[width=14cm]{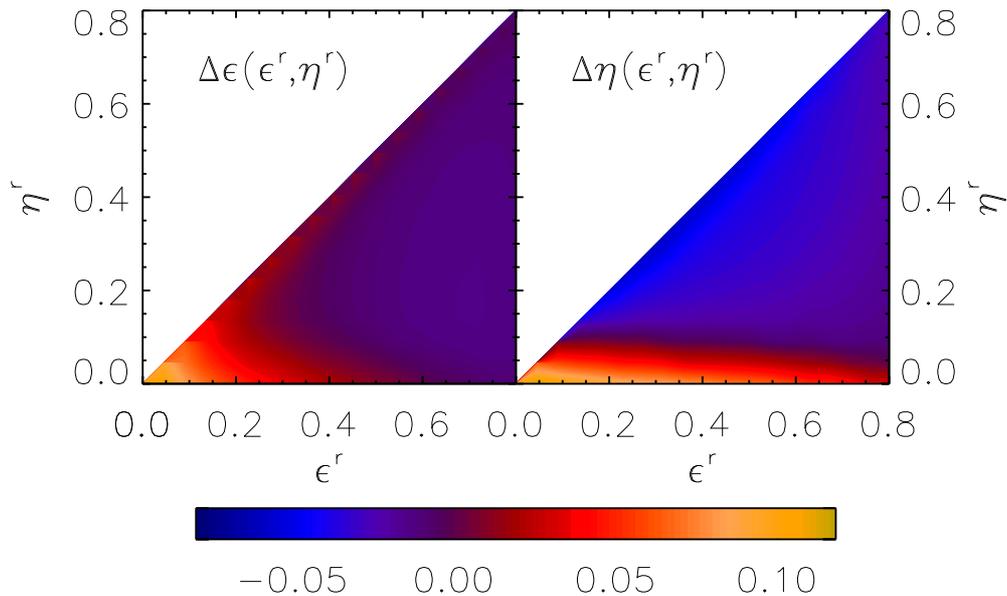}
}%
\caption{%
We show the effect of redshift space distortion on
real space void ellipticities, $\epsilon^r$ and $\eta^r$.
For a given set of ellipticities in real space,
we plot the change in ellipticities,
$\Delta\epsilon(\epsilon^r,\eta^r)
\equiv\epsilon^s(\epsilon^r,\eta^r)-\epsilon^r$ ({\it left}) and
$\Delta\eta(\epsilon^r,\eta^r)\equiv\eta^s(\epsilon^r,\eta^r)-\eta^r$
({\it right}).
Here, we have assumed a top-hat density profile with $\bar{\delta}=-0.9$
and $f=0.5$.
}%
\label{eps_contour}
\end{center}
\end{figure*}
In figure \ref{eps_contour}, we show
$\Delta\epsilon(\epsilon^r,\eta^r)$ ({\it left}) and
$\Delta\eta(\epsilon^r,\eta^r)$ ({\it right})
for $\bar{\delta}=-0.9$ and $f=0.5$.
As expected, a spherical void in the real space becomes
a prolate spheroid (i.e., $\epsilon=\eta\ne 0$) in the redshift space
with its ellipticity given by eq. (\ref{eq:eps_eta_sphere}).
Also, we see a clear preference for positive $\Delta\epsilon$
for any triaxial void, especially at a small $\epsilon^r$.
As for $\Delta\eta$, spheroidal voids tend to be triaxial voids
in the redshift space. From both $\Delta\epsilon$ and $\Delta\eta$,
we find that an oblate void ($T<0.5$) in real space tends
to be a more prolate shape ($T>0.5$) in the redshift space,
and we have more prolate voids in redshift space ellipticity PDF.

\begin{figure*}[t]
\begin{center}
\rotatebox{0}{%
  \includegraphics[width=10cm]{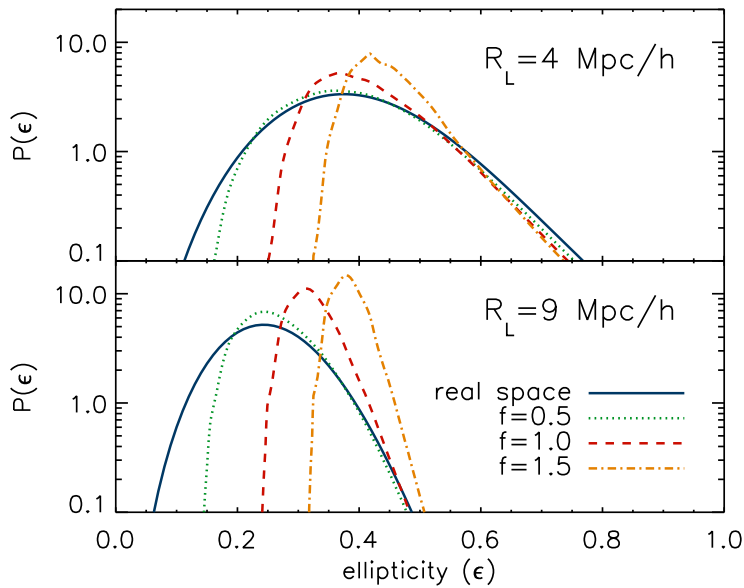}
}%
\caption{%
We compare the ellipticity PDF of real space and redshift space for
$R_L=4h^{-1}{\rm Mpc}$ ({\it top}) and $R_L=9h^{-1}{\rm Mpc}$ ({\it bottom}).
Here, we use $b_L=1$ and $\bar{\delta}=-0.9$, while varying the linear growth
rate: $f=0.5$ (dotted), $1.0$ (dashed) and $1.5$ (dot-dashed).
We also show the real space ellipticity PDF (solid) as a reference.
}%
\label{pdf1d_red}
\end{center}
\end{figure*}

In figure \ref{pdf1d_red}, we compare the ellipticity PDF of
real space and redshift space for $R_L=4h^{-1}{\rm Mpc}$
({\it top}) and $R_L=9h^{-1}{\rm Mpc}$ ({\it bottom}).
Here, we use $b_L=1$ and $\bar{\delta}=-0.9$, while varying the
linear growth rate, $f$, from $0.5$ to $1.5$.
We show the redshift space ellipticity PDF
together with that of real space as a reference.
Since the expansion rate of the void is set by the
matter density distribution inside the void,
in linear theory, it is proportional to the linear growth rate.
As the effect of the redshift space distortion on the
ellipticity PDF is more prominent for a larger $f(z)$ and smaller $\epsilon^r$,
the redshift space ellipticity PDF becomes
more skewed in shape with the mean ellipticity being higher than that of
real space, $\bar{\epsilon}^s>\bar{\epsilon}^r$.
Also note that the left tails of each redshift space ellipticity PDF
approximately show the redshift space ellipticity, $\epsilon^s$, of
a spherical void, $\epsilon^r=0$, and thus, we have
\begin{eqnarray}
\epsilon^{\rm left}\sim\epsilon^s(\epsilon^r=0)=1-\frac{1}{1-\frac{f\bar{\delta}}{3}}.
\end{eqnarray}
Therefore, in principle, with only the location of the left tail of PDF
determined, we can put a constraint
on $f\bar{\delta}$. This is a great benefit of using morphological
information of structure, which is directly distorted in the redshift space.
Although what we can actually measure is $f\bar{\delta}=\beta\bar{\delta}_v$,
where $\beta\equiv f/b_L$, the entire shape of ellipticity PDF can
lift the degeneracy between $f$ and $b_L$.
We can easily see this in the figure \ref{pdf2d_bias},
where the increase in $b_L$ decreases the mean ellipticity,
$\bar{\epsilon^r}$, shifting the right tails of PDF toward lower $\epsilon^r$,
while the left tails are almost unaffected.
Unlike a galaxy power spectrum, $P(k)$, which gives a constraint on the
linear growth rate that is degenerate with the linear bias,
$\beta\equiv f/b_L$, a void ellipticity PDF has a potential to constrain
the linear growth rate separately from the other cosmological parameters.
In other words, for a given galaxy survey, we have the situation
in which we ``{\it buy one, get one free}''.

\subsection{Void With General Radial Density Profile}
So far, we have derived the redshift space ellipticity PDF for
the top-hat density contrast, $\delta(\mathbf{x})$;
however, the realistic radial profile of void density
contrast does not necessarily follow the top-hat shape.
The main parameter controlling the shape of the redshift space
ellipticity PDF is $\kappa(\mathbf{x}^r)$ as in eq. (\ref{eq:kappa}),
and its value is given by $\kappa(\mathbf{x}^r)=A(\mathbf{x}^r)f\bar{\delta}$,
where $A(\mathbf{x}^r)$ is solely determined by the density profile
of a void (e.g., a void with a spherical top hat density profile
has $A(\mathbf{x}^r)=-1/3$).

As an example, we parametrize the radial density profile of the void
following the accumulated density profile of voids from the N-body simulation of
\citet{colberg/etal:2005},
$\rho(<r)/\rho(r_{\rm eff})\propto \exp[(r/r_{\rm eff})^{\alpha}]$,
where the dentity profile at a given radius, $\rho(r)$, is given as follows
\begin{eqnarray}
\rho(r)\propto \exp\left[\left(\frac{r}{r_{\rm eff}}\right)^{\alpha}\right]
\left[1+\frac{\alpha}{3}\left(\frac{r}{r_{\rm eff}}\right)^{\alpha}\right].
\end{eqnarray}

\begin{figure*}[t]
\begin{center}
\rotatebox{0}{%
  \includegraphics[width=10cm]{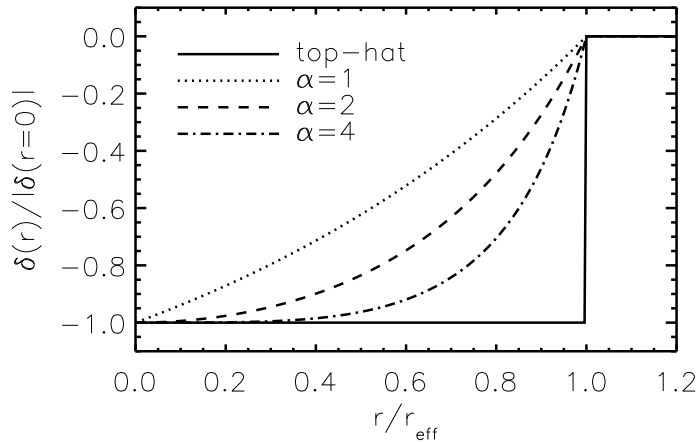}
}%
\caption{%
We show profiles of the density contrast, $\delta(r)$, for top-hat (solid) and
exponential profiles of eq. (\ref{eq:exponential}) with $\alpha=1$ (dotted),
$2$ (dashed) and $4$ (dot-dashed). Here, we assumed that the density at
the effective radius, $r_{\rm eff}$, becomes that of the cosmic mean,
$\rho(r_{\rm eff})=\bar{\rho}$, or, $\delta(r_{\rm eff})=0$.
}%
\label{density_profile}
\end{center}
\end{figure*}

\begin{figure*}[t]
\begin{center}
\rotatebox{0}{%
  \includegraphics[width=10cm]{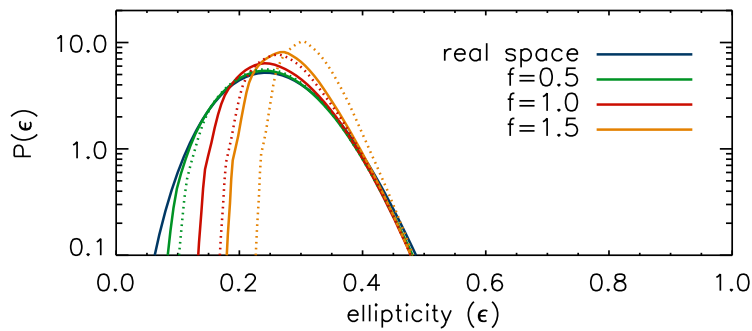}
}%
\caption{%
We show the redshift space ellipticity PDF for $f=0.5$ (green),
$1.0$ (red) and $1.5$ (orange) together with real space ellipticity PDF
(blue). Here, we use $R_L=9h^{-1}{\rm Mpc}$. The radial density profile
of density contrast is set to be either a top-hat (solid), $\delta(r<R_{\rm void})=\bar{\delta}$,
or an exponential (dotted) as in eq. (\ref{eq:exponential}), with
$r_{\rm eff}=R_{\rm void}$ and $\alpha=2$.
We use $\delta_0=-0.40$ and $-0.82$ for the top-hat and
the exponential density profile, respectively, in order to keep the
same mean density contrasts, $\bar{\delta}=-0.4$, for different density profiles.
Here, we see degeneracy between the shape of the potential and
the linear growth rate, $f$.
}%
\label{pdf1d_red_shape}
\end{center}
\end{figure*}

Here, instead of using the original form of \citet{colberg/etal:2005},
we use an exponential density profile,
\begin{eqnarray}
\delta(r)=\delta_0\left(\frac{e^{[(r/r_{\rm eff})^{\alpha}-1]}-1}{e^{-1}-1}\right),
\label{eq:exponential}
\end{eqnarray}
where we forced $\rho(r_{\rm eff})=\bar{\rho}$, at the effective radius,
$r_{\rm eff}$, (i.e., $\delta(r_{\rm eff})=0$), and $\delta_0\equiv\delta(r=0)$.
Figure \ref{density_profile} shows the profiles of the density contrast, $\delta(r)$,
for top-hat (solid) and
exponential profiles of eq. (\ref{eq:exponential}) with $\alpha=1$ (dotted),
$2$ (dashed) and $4$ (dot-dashed).
Note that for a top-hat density contrast, we have $\bar{\delta}=\delta_0$,
while for the exponential profiles, we have
\begin{eqnarray}
\bar{\delta}=\frac{\int_0^{r_{\rm eff}} r^2 \delta(r)dr}{\int_0^{r_{\rm eff}} r^2 dr},
\end{eqnarray}
where $\delta (r)$ is given as eq. (\ref{eq:exponential}).
In figure \ref{pdf1d_red_shape}, we show the redshift space ellipticity
PDF for $f=0.5$ (green), $1.0$ (red) and $1.5$ (orange) together with the
real space ellipticity PDF (blue).
Here, we use $R_L=9h^{-1}{\rm Mpc}$ and the radial density profile
of the density contrast is set to be either a top-hat (solid),
$\delta(r<R_{\rm void})=\bar{\delta}$,
or an exponential (dotted) given by eq. (\ref{eq:exponential}) with
$r_{\rm eff}=R_{\rm void}$ and $\alpha=2$.
We use $\delta_0=-0.40$ and $-0.82$ for the top-hat density profile and
the exponential profile, respectively, in order to keep the
same mean density contrasts, $\bar{\delta}=-0.4$, for different density profiles.
Here, we clearly see degeneracy between the shape of the potential and
the linear growth rate, $f$.
Therefore, a better understanding of the density profile of a void
is required when constraining the linear growth rate from
the redshift space ellipticity PDF.

\section{Effect of Void Finding and Poisson Noise on the Void Ellipticity PDF}
\label{sec:poisson_noise}
In the previous sections, we have shown how to incorporate the galaxy bias
and redshift space distortion into the calculation of the ellipticity PDF.
In this section, we investigate the biases and errors that arise when using the inertia tensor,
$I_{ij}$, as an estimate of the void ellipticity.
Since a void is a sparse region in a galaxy distribution,
the number of galaxies inside a void, $N_{\rm vg}$, is limited.
Therefore the error in the ellipticity calculated from the
inertia tensor of the void galaxy distribution is dominated by the Poisson statistics.

Let us define two stochastic variables, $x$ and $y$, as follows,
\begin{eqnarray}
x\equiv \mu_x+\Delta x,\\
y\equiv \mu_y+\Delta y,
\end{eqnarray}
where $\mu_x$ and $\mu_y$ are the means of $x$ and $y$, respectively,
and $\left<\Delta x\right>=\left<\Delta y\right>=0$.
We have variances and covariance defined as,
\begin{eqnarray}
\sigma_x^2&\equiv&\left<(x-\mu_x)^2\right>=\left<(\Delta x)^2\right>,\\
\sigma_y^2&\equiv&\left<(y-\mu_y)^2\right>=\left<(\Delta y)^2\right>,\\
\sigma_{xy}&\equiv&\left<(x-\mu_x)(y-\mu_y)\right>=\left<\Delta x\Delta y\right>.
\end{eqnarray}
A mean of a rational is given by the
following form,
\begin{eqnarray}
\mu_{x/y}=\left<\frac{\mu_x+\Delta x}{\mu_y+\Delta y}\right>.
\end{eqnarray}
Expanding $\Delta y$ up to the second order, we have
\begin{eqnarray}
  \mu_{x/y}&\simeq&\frac{\mu_x}{\mu_y}\left<\left[1+\frac{\Delta x}{\mu_x}\right]
  \left[1-\frac{\Delta y}{\mu_y}+\left(\frac{\Delta y}{\mu_y}\right)^2\right]\right>
  \nonumber\\
  &\simeq&\frac{\mu_x}{\mu_y}
  \left(
  1+\frac{\sigma_y^2}{\mu_y^2}
  -\frac{\sigma_{xy}}{\mu_x\mu_y}
  \right).
\end{eqnarray}
Similarly, variance of the rational is
\begin{eqnarray}
  \sigma_{x/y}^2&=&\left<\left(\frac{\mu_x+\Delta x}{\mu_y+\Delta y}\right)^2\right>-\mu_{x/y}^2
  \nonumber\\
  &\simeq&\frac{\mu_x^2}{\mu_y^2}
  \left(
  \frac{\sigma_x^2}{\mu_x^2}
  +\frac{\sigma_y^2}{\mu_y^2}
  -\frac{2\sigma_{xy}}{\mu_x\mu_y}
  \right).
\end{eqnarray}
Now, ellipticity, $\epsilon$, is defined as $\epsilon=1-\frac{p_3}{p_1}=1-\sqrt{\frac{I_3}{I_1}}$,
where $p_1\ge p_2\ge p_3$ are the three principal axes of the ellipsoid,
and $I_1\ge I_2\ge I_3$ are the eigenvalues of the inertia tensor $I_{ij}$.
From $p_i^2\propto I_i$, we have
\begin{eqnarray}
\frac{\Delta p_i}{\mu_{p_i}}=\frac12\frac{\Delta I_i}{\mu_{I_i}},
\end{eqnarray}
and therefore, we have
\begin{eqnarray}
\mu_{\epsilon}&=&1-\mu_{p_3/p_1}\simeq 1-\frac{\mu_{p_3}}{\mu_{p_1}}
\left(1+\frac{\sigma_{p_1}^2}{\mu_{p_1}^2}-\frac{\sigma_{p_1p_3}}{\mu_{p_1}\mu_{p_3}}\right)
\nonumber\\
&=&1-\sqrt{\frac{\mu_{I_3}}{\mu_{I_1}}}
\left(
1+\frac14\frac{\sigma_{I_1}^2}{\mu_{I_1}^2}
-\frac14\frac{\sigma_{I_1I_3}}{\mu_{I_1}\mu_{I_3}}
\right),
\label{eq:mean_epsilon}
\end{eqnarray}
and
\begin{eqnarray}
\sigma_{\epsilon}^2=\sigma_{p_3/p_1}^2&\simeq&\frac{\mu_{p_3}^2}{\mu_{p_1}^2}
\left(
\frac{\sigma_{p_1}^2}{\mu_{p_1}^2}+\frac{\sigma_{p_3}^2}{\mu_{p_3}^2}
-\frac{2\sigma_{p_1p_3}}{\mu_{p_1}\mu_{p_3}}
\right)
\nonumber\\
&=&\frac14\frac{\mu_{I_3}^2}{\mu_{I_1}^2}
\left(
\frac{\sigma_{I_1}^2}{\mu_{I_1}^2}+\frac{\sigma_{I_3}^2}{\mu_{I_3}^2}
-\frac{2\sigma_{I_1I_3}}{\mu_{I_1}\mu_{I_3}}
\right).
\label{eq:stdev_epsilon}
\end{eqnarray}
From eq. (\ref{eq:inertia_real}), we have
\begin{eqnarray}
I_{ij}\equiv\frac1{N_{\rm vg}}\sum^{N_{\rm vg}}_{\alpha=1}x_{i,\alpha}x_{j,\alpha},
\end{eqnarray}
and in the continuous density field, we have
\begin{eqnarray}
I_{ij}&\equiv&\int_{V_{void}}d^3x~ n({\mathbf{x}})x_ix_j
\nonumber\\
&=&\int_{V_{void}}d^3x~ \left[\bar{n}+\delta n({\mathbf{x}})\right]x_ix_j
=\mu_{I_{ij}}+\Delta I_{ij},
\end{eqnarray}
where
\begin{eqnarray}
\mu_{I_{ij}}=\int_{V_{void}}d^3x~ \bar{n}x_ix_j,\\
\Delta I_{ij}=\int_{V_{void}}d^3x~\delta n({\mathbf{x}})x_ix_j.
\end{eqnarray}
Also, the variance of the inertia tensor is given as
\begin{eqnarray}
\sigma_{I_{ij}}^2=\left<(\Delta I_{ij})^2\right>=\left<I_{ij}^2\right>-\mu_{I_{ij}}^2.
\end{eqnarray}

\subsection{Void With a Well-defined Boundary: $\left<p_i\right>\to P_i$}
Let us calculate the bias in the mean and the variance of the ellipticity
calculated from a rectangular void with a diagonalized inertia tensor of
$I_1\equiv I_{11}$, $I_2\equiv I_{22}$ and $I_3\equiv I_{33}$. The number of void galaxies is
$N_{\rm vg}$.
Here, we assume an idealized setup, where the boundary shape of a void is well
defined (i.e., $\left<p_i\right>\to P_i$) to see whether the estimator of an output ellipticity based on the
inertia tensor yields unbiased result.
For more realistic estimate on the bias in mean and the variance of output ellipticities,
see \S~\ref{sec:undefined_boundary}.

For a rectangular void with volume,
$V_{void}=8P_1P_2P_3$ (i.e., $x_i\in [-P_i,P_i]$ $\forall~i$),
where $P_i$ is a well-defined length of the principal axis of the rectangular void, we have
\begin{eqnarray}
\mu_{I_{i}}=\int_{V_{void}}d^3x~ \bar{n}x_i^2=\frac13\bar{n}V_{void}P_i^2=\frac13N_{\rm vg}P_i^2,
\label{eq:mu_rect}
\end{eqnarray}
and
\begin{eqnarray}
\sigma_{I_{i}}^2&=&\int_{V_{void}}\int_{V_{void}'}d^3xd^3x'~
\left<n({\mathbf{x}})n({\mathbf{x}'})\right>x_i^2x_i'^2
-\mu_{I_i}^2
\nonumber\\
&=&\int_{V_{void}}\int_{V_{void}'}d^3xd^3x'~
\bar{n}\delta_D(\mathbf{x}-\mathbf{x}')x_i^2x_i'^2-\mu_{I_i}^2
\nonumber\\
&=&\bar{n}\int_{V_{void}}d^3x~x_i^4-\mu_{I_i}^2
=\frac15\bar{n}V_{void}P_i^4-\mu_{I_i}^2
\nonumber\\
&=&\frac15N_{\rm vg}P_i^4-\frac19N_{\rm vg}^2P_i^4,
\label{eq:sig_rect}
\end{eqnarray}
where we have assumed that the distribution of void galaxies follows the Poisson statistics with
$\left<n(\mathbf{x})\right>=\bar{n}$ and
$\left<n(\mathbf{x})n(\mathbf{x}')\right>=\bar{n}\delta_D(\mathbf{x}-\mathbf{x}')$.
Similarly, we have
\begin{eqnarray}
\sigma_{I_{i}I_{j}}&=&\bar{n}\int_{V_{void}}d^3x~x_i^2x_j^2-\mu_{I_i}\mu_{I_j}
\nonumber\\
&=&\frac19N_{\rm vg}P_i^2P_j^2-\frac19N_{\rm vg}^2P_i^2P_j^2,
\end{eqnarray}
Therefore, Eqs(\ref{eq:mean_epsilon}) and (\ref{eq:stdev_epsilon}) are re-written as follows,
\begin{eqnarray}
\mu_{\epsilon}&=&1-\frac{P_3}{P_1}-\frac{P_3}{P_1}\frac1{5N_{\rm vg}},
\label{eq:mean_epsilon_rect}\\
\sigma_{\epsilon}^2&=&\frac{P_3^2}{P_1^2}\frac{2}{5N_{\rm vg}}.
\label{eq:stdev_epsilon_rect}
\end{eqnarray}
We see that for a given rectangular void with $N_{void}=20$, we can accurately extract its ellipticity
from void galaxies to $\sim 1\%$, but its standard deviation remains significant
(i.e., $\sigma_{\epsilon}\sim 0.14$ for a void with zero ellipticity).
As is expected, the variance of ellipticity, $\sigma_{\epsilon}^2$,
scales as inversely proportional to the number of void galaxies, $N_{\rm vg}$.
Note that the above result is derived in the limit of
$\left<p_i\right>\to P_i$.

\subsection{Void With an Undefined Boundary: $\left<p_i\right>\ne P_i$}
\label{sec:undefined_boundary}
With a small number of field galaxies randomly distributed
inside the well-defined boundary, $P_i$, we generally have
$\left<p_i\right>\ne P_i$.
For example, let us imagine a spherical boundary, where $P_1=P_2=P_3$
and $\epsilon_{in}=1-\frac{P_3}{P_1}=0$, with $N_{\rm vg}=4$ field galaxies.
This spherical boundary represents the shape of the underlying tidal field,
and we sample this by 4 galaxies.
No matter how we distribute those four galaxies, unless the four galaxies
align at each of four apexes of an equilateral tetrahedron, calibrated
ellipticity cannot be zero, $\epsilon_{out}\ne \epsilon_{in}$.
As a result, we have a significant deviation of $\left<\epsilon_{out}\right>$
from its boundary ellipticity, $\epsilon_{in}\equiv 1-\frac{P_3}{P_1}$,
and a correspondingly large standard deviation.

As we increase $N_{\rm vg}$, those randomly scattered particles fill in the
bounded space more densely, and $\left<\epsilon_{out}\right>$ gradually approaches
the boundary ellipticity, $\epsilon_{in}$.
This uncertainty in $\epsilon_{out}$ due to small $N_{\rm vg}$
introduces an additional bias and scatter to the ones given by
eqs. (\ref{eq:mean_epsilon_rect}) and (\ref{eq:stdev_epsilon_rect}).

For the case of elliptical voids, the analytic prediction of the mean and variance
is non-trivial; thus, we have performed numerical experiments.
First, we create an ellipsoidal boundary with a fixed size and ellipticity
(i.e., $P_1$, $P_2$ and $P_3$),
and then fill this bounded volume by $N_{\rm vg}$ random particles.
For each realization, we calculate the inertia tensor
and its eigenvalues ($p_1$, $p_2$ and $p_3$),
where the output ellipticity is defined as
\begin{eqnarray}
\epsilon_{out}\equiv 1-\frac{p_3}{p_1},\\
\eta_{out}\equiv 1-\frac{p_2}{p_1},
\end{eqnarray}
while the input ellipticity is defined as
\begin{eqnarray}
\epsilon_{in}\equiv 1-\frac{P_3}{P_1},\\
\eta_{in}\equiv 1-\frac{P_2}{P_1}.
\end{eqnarray}
Finally, we have a bias in the mean and variance of the output ellipticity,
$\epsilon^{out}$ as follows
\begin{eqnarray}
\left<\epsilon_{out}\right>-\epsilon_{in}=\frac{P_3}{P_1}-\left<\frac{p_3}{p_1}\right>,\\
\sigma^2_{\epsilon_{out}}=\left<\epsilon_{out}^2\right>-\left<\epsilon_{out}\right>^2.
\end{eqnarray}

\begin{figure*}[t]
\begin{center}
\rotatebox{0}{%
  \includegraphics[width=12cm]{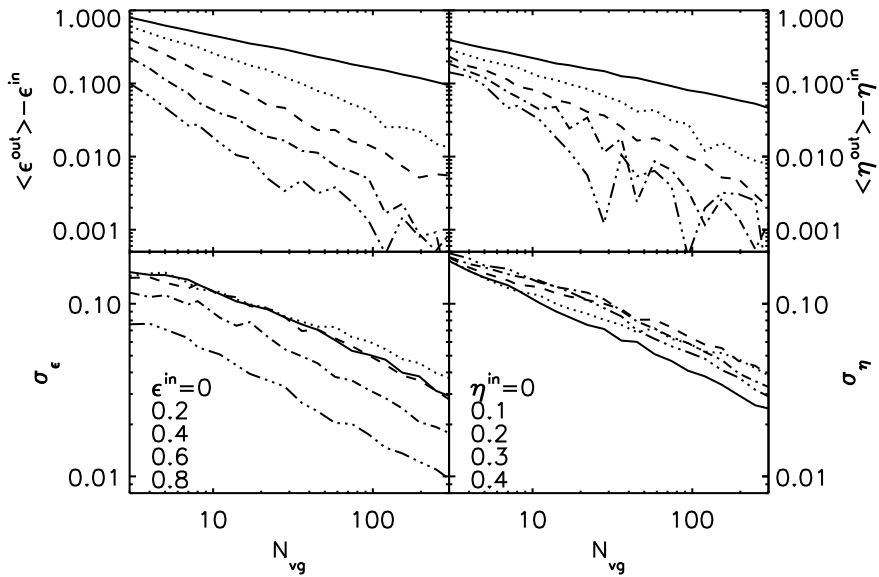}
}%
\caption{%
Bias and variance in the measured ellipticities. We place $N_{\rm vg}$ particles
within a boundary of a given set of size and input ellipticities,
$\epsilon^{\rm in}$ and $\eta^{\rm in}$, and
calculate the mean ellipticities, $\left<\epsilon^{\rm out}\right>$ and $\left<\eta^{\rm out}\right>$,
and their standard deviations $\sigma_{\epsilon}$ and $\sigma_{\eta}$,
from 500 realizations for each $N_{\rm vg}$ and input ellipticity.
Top figures show the biases in finding the input ellipticities
(i.e., true underlying ellipticities of a tidal field),
$\left<\epsilon^{\rm out}\right>-\epsilon^{\rm in}$ and
$\left<\eta^{\rm out}\right>-\eta^{\rm in}$,
and bottom figures show the standard deviations,
$\sigma_{\epsilon}$ and $\sigma_{\eta}$.
}%
\label{response}
\end{center}
\end{figure*}

\begin{figure*}[t]
\begin{center}
  \rotatebox{0}{%
    \includegraphics[width=12cm]{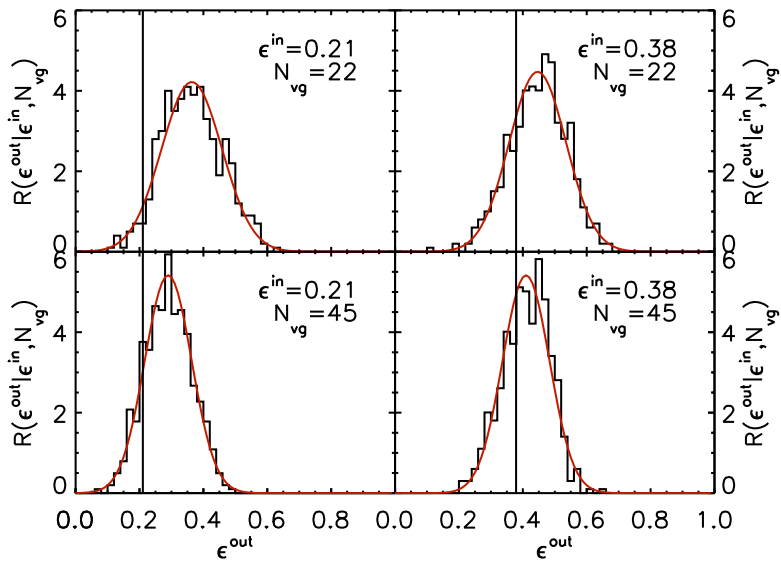}
  }%
  \caption{%
    We show the normalized histograms of output ellipticities, $\epsilon_{\rm out}$
    for a fixed input ellipticity, $\epsilon_{\rm in}$, and a number of
    void galaxies, $N_{\rm vg}$. We find that the Gaussian profile
    of eq. (\ref{eq:response})
    with mean, $\left<\epsilon_{\rm out}\right>$, and variance,
    $\sigma_{\epsilon_{\rm out}}$, fits the measured histograms (red lines) well.
    Also, we show the locations of the input ellipticity, $\epsilon_{\rm in}$,
    with vertical lines.
  }%
  \label{response_func}
\end{center}
\end{figure*}

\begin{figure*}[t]
\begin{center}
\rotatebox{0}{%
  \includegraphics[width=12cm]{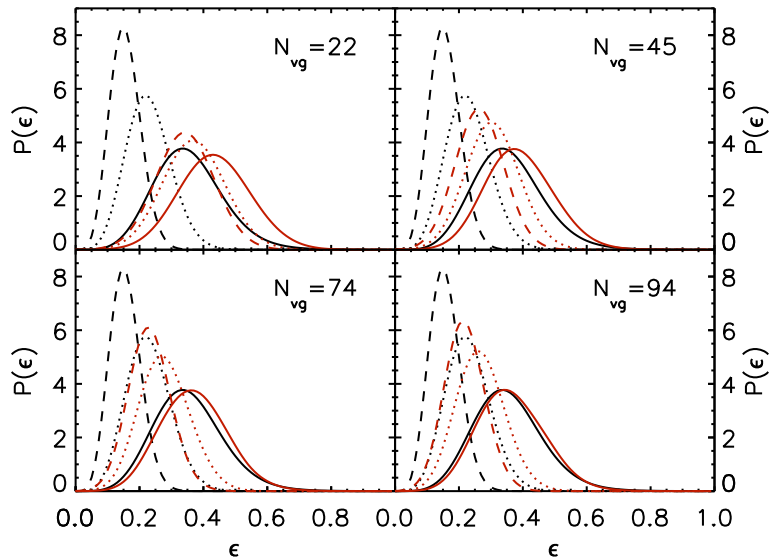}
}%
\caption{%
  We show the real space ellipticity PDFs without Poisson noise (black lines)
  at $z=0$ (solid), $1$ (dotted) and $2$ (dashed).
  We then convolve each noise-free PDF with the response function given by eq. (\ref{eq:response}),
  and show the resulting PDFs for four different $N_{\rm vg}$ (red lines).
}%
\label{pdf_response}
\end{center}
\end{figure*}

Figure \ref{response} shows the result of 500 realizations for a given set of
$\epsilon^{\rm in}$ and $N_{\rm vg}$.
The top figures show the bias in the mean of the ellipticities
$\epsilon$ ({\it left}) and $\eta$ ({\it right}), and the bottom figures show
the standard deviations of $\epsilon$ ({\it left}) and $\eta$ ({\it right}).
We see that both bias in the mean and variance of $\epsilon_{\rm out}$
are large enough to change the shape of ellipticity PDF significantly for
$N_{\rm vg}\lesssim 100$.
In order to correct the ellipticity PDF, we define a response
function, $R(\epsilon_{\rm out}|\epsilon_{\rm in},N_{\rm vg})$, such that
\begin{eqnarray}
&&p(\epsilon_{\rm out}|\sigma_{R_L},\delta,N_{\rm vg})
\nonumber\\
&&=\int d\epsilon_{\rm in} R(\epsilon_{\rm out}|\epsilon_{\rm in},N_{\rm vg})
p(\epsilon_{\rm in}|\sigma_{R_L},\delta).
\end{eqnarray}
In figure \ref{response_func}, we show the normalized histograms for
the number of voids with output ellipticity, $\epsilon_{\rm out}$,
from the numerical simulation with a given $N_{\rm vg}$ and $\epsilon_{\rm in}$.
We also show Gaussian distributions (red lines) with the mean $\left<\epsilon_{\rm out}\right>$ and
variance $\sigma^2_{\epsilon_{\rm out}}$,
\begin{eqnarray}
R(\epsilon_{\rm out}|\epsilon_{\rm in},N_{\rm vg})=\frac1{\sqrt{2\pi\sigma^2_{\epsilon_{\rm out}}}}
\exp\left[-\frac{(\epsilon_{\rm out}-\left<\epsilon_{\rm out}\right>)^2}{2\sigma^2_{\epsilon_{\rm out}}}
\right],
\label{eq:response}
\nonumber\\
\end{eqnarray}
where both mean and variance are functions of the input ellipticity and the number of void galaxies
(i.e., $\left<\epsilon_{\rm out}\right>=\left<\epsilon_{\rm out}(\epsilon_{\rm in},N_{\rm vg})\right>$
and $\sigma_{\epsilon_{\rm out}}=\sigma_{\epsilon_{\rm out}}(\epsilon_{\rm in},N_{\rm vg})$).
In figure \ref{pdf_response}, we show the ellipticity PDF without the correction
for the response function (black lines) as well as the corrected ellipticity PDF (red lines)
for four different redshifts, $z=0$ (solid), $1$ (dotted), $2$ (dashed)
and $3$ (dot-dashed) with different values of $N_{\rm vg}$.
Here, we use $R_L=4{\rm Mpc^{-1}}$ and $\delta_m=-0.9$ to calculate
the original PDFs at four different redshifts.
We see that the shapes of PDF with higher redshifts are more susceptible to
the small $N_{\rm vg}$ effect.
For a typical void studied in PL07 using the Millennium Run catalog at $z=0$,
$\bar{N}_{\rm vg}$ is roughly between 20 to 60,
$R_L$ ranges from $4$ to $6.4h^{-1}{\rm Mpc}$, and $\delta_v\sim -0.9$.
We clearly see that voids with small initial ellipticity
$\epsilon_{\rm in}$ are most vulnerable to the Poisson noise,
and with $N_{\rm vg}\sim 10$, the original shape of ellipticity PDF is
almost washed out to leave only the distribution of the Poisson noise.
On the other hand, the ellipticity PDF at low redshift ($z=0$)
maintains its original shape with the number of void galaxies as small as
$N_{\rm vg}\lesssim 45$.
In short, a shape with small ellipticity is hard to trace with
a small number of tracers. For example, if we have a sphere of radius
$R=10h^{-1}{\rm Mpc}$ in the simulation box with a resolution of
$1h^{-1}{\rm Mpc}$, the possible number of configurations of randomly scattered
tracing particles with $N_{\rm vg}=4$ will be
$_{4187}C_{4}\sim 1.3\times10^{13}$, while the possible number of configurations
of an equilateral tetrahedron (i.e., $\epsilon=0$) is vanishingly small.
Also, the scatter of the output ellipticity $\sigma_{\epsilon}$ will not
center around $\epsilon_{\rm in}$ for a small ellipticity as
$0\le\epsilon\le 1$, and as a result, $\left<\epsilon_{\rm out}\right>$
converges to $\epsilon_{\rm in}$ only slowly as we increase $N_{\rm vg}$.
On the other hand, the ellipticity of a highly elongated shape
can be well traced even with a limited number of tracers.

For $N_{\rm vg}\lesssim 100$, we use a fitting function derived from figure \ref{response},
\begin{eqnarray}
\left<\epsilon_{out}\right>-\epsilon_{in}=\alpha_1(\epsilon_{in}) N_{\rm vg}^{-\beta_1(\epsilon_{in})}
\label{eq:fit_func_bias}
\\
\sigma_{\epsilon_{\rm out}}=\alpha_2(\epsilon_{in}) N_{\rm vg}^{-\beta_2(\epsilon_{in})},
\label{eq:fit_func_stdv}
\end{eqnarray}
where $\alpha_1$, $\alpha_2$, $\beta_1$, and $\beta_2$ are given as
polynomials to the third order, $p_0+p_1\epsilon_{\rm in}+p_2\epsilon_{\rm in}^2+p_3\epsilon_{\rm in}^3$,
and we show the best-fit coefficients, $p_0$, $p_1$, $p_2$ and $p_3$
in table \ref{tb:coeff}.
\begin{table*}
\begin{center}
\begin{tabular}{|c|c|c|c|c|}
\hline
 & $p_0$ & $p_1$ & $p_2$ & $p_3$ \\

\hline
$\alpha_1$ & 1.282 & 2.384 & -7.280 & 3.587 \\
\hline
$\alpha_2$ & 0.2636 & -0.3103 & 0.9557 & -0.9404 \\
\hline
$\beta_1$  & 0.4500 & 1.708 &  -0.8460 & -0.02668 \\
\hline
$\beta_2$  & 0.3650 & -0.6260 & 1.680 & -0.9296 \\
\hline
\end{tabular}
\caption{%
We show the best-fit coefficients, $p_0$, $p_1$, $p_2$ and $p_3$
for $\alpha_1$, $\alpha_2$, $\beta_1$, and $\beta_2$ of
eqs. (\ref{eq:fit_func_bias}) and (\ref{eq:fit_func_stdv}),
where $\alpha_1$, $\alpha_2$, $\beta_1$, and $\beta_2$
are given as polynomials to the third order,
$p_0+p_1\epsilon_{\rm in}+p_2\epsilon_{\rm in}^2+p_3\epsilon_{\rm in}^3$.
}%
\label{tb:coeff}
\end{center}
\end{table*}

\section{Voids from N-body Simulation}
\label{sec:n-body}
To test the analytic predictions made in the previous sections,
we use a catalog of galaxies from the Millennium simulation
\citep{croton/etal:2006,delucia/blaizot:2007}.
In a comoving box of $L_{{\rm box}}=500~h^{-1}{\rm Mpc}$,
there are $26690265$, $26359329$, and $23885840$ galaxies in the simulation
at $z=0$, $1$, and $2$, respectively.
For each redshift, we select a subset of galaxies whose halo masses are above $M_h=2.6\times 10^{12}$,
$2.6\times 10^{12}$, and $8.6\times 10^{11}h^{-1}M_{\odot}$.
We find $182081$, $121454$, and $261471$ such halos for $z=0$, $1$, and $2$, respectively.

Here, we use the void finding algorithm of FN09 to identify voids from the subsets of galaxies.
Their algorithm first divides a given set of galaxies into wall galaxies, which define
boundaries of holes, and void galaxies, which are allowed to be within holes
by calculating the distances to the third nearest galaxies, $D_3$.
When a galaxy has the third nearest distance, $D_3$, smaller than the
user-defined threshold distance, $R_3>D_3$, the galaxy is labeled as a wall galaxy,
and when $R_3<D_3$, the galaxy is labeled as a void galaxy.
Here, $R_3\equiv \left<D_3\right>+\lambda \sigma_{D_3}$, where $\left<D_3\right>$ and
$\sigma_{D_3}$ are the mean third distance and its standard deviation, respectively,
and $\lambda$ is a user-defined parameter.
We use the value of $\lambda=2$ recommended by FN09.
Then, a hole is defined as a sphere with maximal radius in a galaxy distribution containing no wall galaxy,
and by definition, a hole can contain void galaxies.
Finally, a void is identified by the subsequent merger of neighboring holes with user-defined
criteria (i.e., a percentage of volume overwraps with the neighboring holes and a minimum size of hole, $\xi$),
where, we use $\xi=1.5R_3$.
As a result, voids identified by the algorithm of FN09 have aspherical shape
with a finite number of void galaxies within.
For each identified void, FN09 found a volume, a size and an ellipticity
of the best-fit ellipsoid by calculating the inertia tensor from randomly scattered
particles within the boundary of the void.

Using a void-finder of FN09, we identify $569$, $327$, and $804$ voids for
$z=0$, $1$, and $2$, respectively.
Here, as we noted earlier, since we are interested in a
shape of a tidal field near the density minima, we have modified the
void finding algorithm of FN09 so that an ellipticity of a void
can be evaluated from the distributions of galaxies inside a
void (i.e., we directly calculate an inertia tensor from void galaxies, not from
randomly scattered particles as in FN09).

With this modification, the extracted ellipticity PDF has
a closer contact with the analytic ellipticity PDF derived from the
tidal field of the local minima of the density field.

\begin{figure*}[t]
\begin{center}
\rotatebox{0}{%
  \includegraphics[width=10cm]{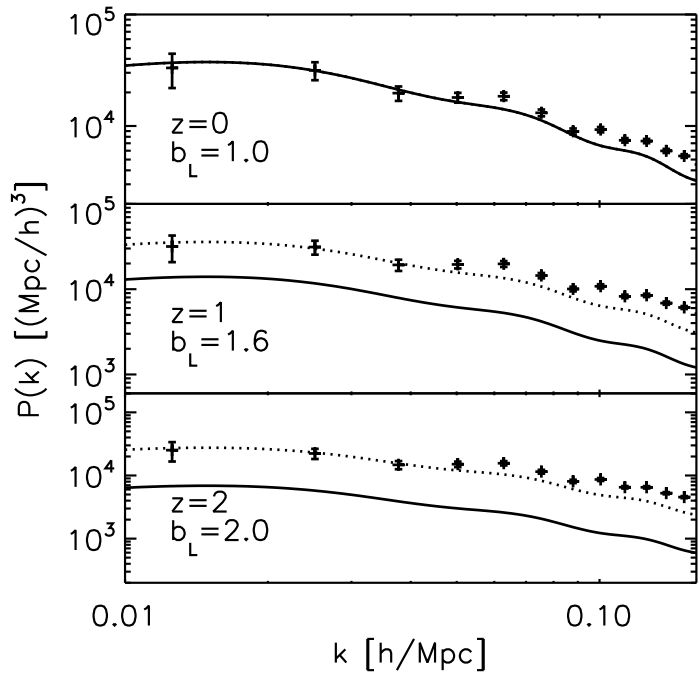}
}%
\caption{%
We show the linear matter power spectra, $P_m(k)$,
from CAMB with the cosmological parameters same as the Millennium simulation
(i.e., $\Omega_m=0.25$, $\Omega_{\Lambda}=0.75$, $n_s=1$ and $\sigma_8=0.9$)
(solid), linear galaxy spectra, $P_g(k)=b_L^2P_m(k)$, (dotted)
and power spectra of the subset of galaxies from the Millennium simulation
(crosses with error bars)
at $z=0$ (top), $1$ (middle) and $2$ (bottom).
}%
\label{pk_bias}
\end{center}
\end{figure*}

We measure the linear galaxy biases by comparing the
galaxy power spectra of input galaxy distributions
against linear matter power spectra from CAMB at each redshift.
In figure \ref{pk_bias}, we show the linear matter power spectra, $P_m(k)$,
from CAMB with the cosmological parameters same as the Millennium simulation
(i.e., $\Omega_m=0.25$, $\Omega_{\Lambda}=0.75$, $n_s=1$ and $\sigma_8=0.9$)
(solid), linear galaxy spectra, $P_g(k)=b_L^2P_m(k)$, (dotted)
and power spectra of the subset of galaxies from the Millennium simulation
(crosses with error bars)
at $z=0$ (top), $1$ (middle) and $2$ (bottom).
With fittings up to $k_{max}=0.03h~{\rm Mpc}^{-1}$
(i.e., using the first two data points of $P_g(k)$),
we find $b_L=1.0$, $1.6$, and $2.0$ for $z=0$, $1$, and $2$, respectively.
Here, non-linearity of the power spectrum is rather small at this scale,
$k<0.03h~{\rm Mpc}^{-1}$.

\begin{figure*}[t]
\begin{center}
\rotatebox{0}{%
  \includegraphics[width=10cm]{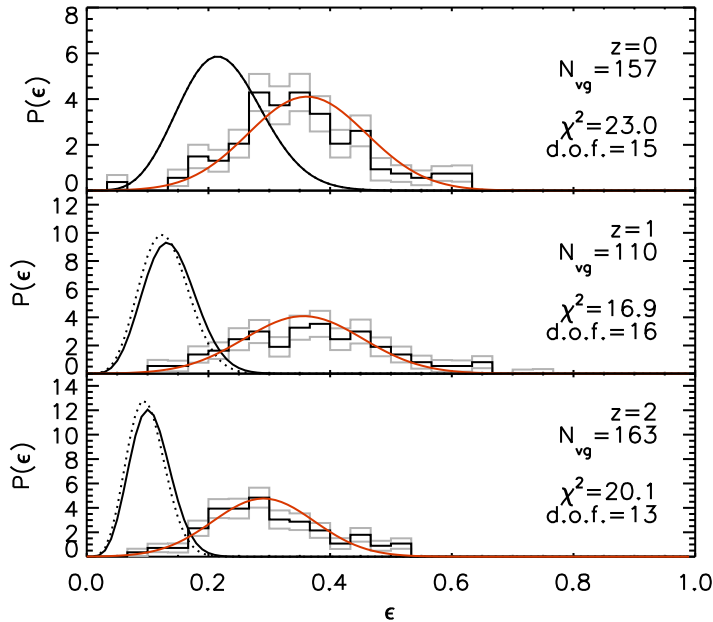}
}%
\caption{%
We show the ellipticity PDF extracted from the Millennium simulation at
$z=0$ (top), $1$ (middle) and $2$ (bottom) together with the
analytic ellipticity PDF corrected for galaxy biases and convolved with the response functions (red lines).
As references, we also show the analytic ellipticity PDF (black lines)
with (dotted black lines) and without galaxy bias (solid black lines).
We also show the $1-\sigma$ range of the histogram, $\sigma_{\epsilon}$,
estimated via the bootstrap method (gray histograms).
For each redshift, the chi-square values are $\chi^2=21.5$,
$17.0$ and $20.0$, and the degree of freedoms are $15$, $16$
and $13$ for $z=0$, $1$ and $2$, respectively.
}%
\label{mill_void}
\end{center}
\end{figure*}

In figure \ref{mill_void}, we show the ellipticity PDF extracted from
the Millennium simulation at $z=0$ (top), $1$ (middle) and $2$ (bottom) together with
the analytic ellipticity PDF corrected for the linear galaxy biases and response functions (red lines).
As references, we also show the analytic ellipticity PDF (black lines)
with (dotted black lines) and without the linear galaxy bias (solid black lines).
We also show the $1$-$\sigma$ range of the histograms
calculated with bootstrapping method (grey).

Here, for each redshift, we demand voids have at least 10 void galaxies.
Table \ref{tb:mill_void} shows redshifts, $z$, the numbers of voids used to draw histograms, $N_{\rm v}$,
the minimum numbers of void galaxies per void, $N_{\rm vg,min}$,
the median numbers of void galaxies per void, $N_{\rm vg,med}$, 
the linear galaxy biases, $b_L$, the mean density contrasts of void, $\bar{\delta}_{\rm v}$,
the mean Eulerian radii of voids, $\bar{R}_E\equiv(\overline{p_1p_2p_3})^{1/3}$,
the mean ellipticities of extracted voids, $\bar{\epsilon}^{\rm obs}$, and those of the analytic PDF,
$\bar{\epsilon}^{\rm theory}$, and the errors in the mean, $\Delta\epsilon^{\rm obs}$.

We find from the figure \ref{mill_void} that the
analytic ellipticity PDFs convolved with the response function (red lines)
show a good agreement with the extracted ellipticity PDFs
from the simulations, with the chi square values of
$\chi^2=23.0$, $16.9$ and $20.1$ with the degrees of freedom of
$15$, $16$ and $13$ for redshifts of $z=0$, $1.078$ and $2.070$, respectively.
We test the null hypothesis that the extracted ellipticity PDF is drawn from
the analytic ellipticity PDF convolved with the response function,
calculating the probability-to-exceed (PTE) the measured $\chi^2$,
\begin{eqnarray}
\alpha=\int^{\infty}_{\chi^2}P_n(x)dx,
\end{eqnarray}
where $P_n(x)$ is a PDF of $\chi^2$-distribution with a $n$-degree of freedom,
\begin{eqnarray}
P_n(x)=\frac{x^{n/2-1}e^{-x^2/2}}{2^{n/2}\Gamma(n/2)}.
\end{eqnarray}
We find that our analytic model is consistent with the void ellipticity PDF
extracted from the Millennium simulation with
${\rm PTE}=0.12$, $0.39$ and $0.094$ for redshifts of
$z=0$, $1.078$ and $2.070$, respectively.

Although we have good fits for all redshifts,
the extracted ellipticity PDFs are entirely dominated by the Poisson
noise due to the small number of void galaxies, $N_{\rm vg}\ll 100$.
To see this, we have performed a null test with a top-hat PDF with
the mean and the width equal to the analytic ellipticity PDF.
After convolving the top-hat PDF with the response functions,
we find an equally good fit with a similar $\chi^2$.
We repeat the null tests for different values of the mean and the width
of the analytic ellipticity PDF.
We find that the extracted ellipticity PDF is well fitted with
an arbitrary shape of PDF, whose mean ellipticity is the same as the one
predicted from the analytic ellipticity PDF, convolved with the response function.

\begin{table*}
\begin{center}
\begin{tabular}{|c|c|c|c|c|c|c|c|c|c|}
\hline
$z$ & $N_{\rm v}$ & $N_{\rm vg,min}$ & $N_{\rm vg,med}$ & $b_L$ & $\delta_v$ & $\bar{R}_E$ &
$\bar{\epsilon}^{\rm obs}$ & $\bar{\epsilon}^{\rm theory}$ & $\Delta\epsilon^{\rm obs}$\\
\hline
0      & 157 & 15 & 19 & 1.0 & -0.66 & 21.8 & 0.343 & 0.360 & 0.008\\
\hline
1.078      & 110 & 10 & 15 & 1.6 & -0.59 & 21.9 & 0.377 & 0.356 & 0.012\\
\hline
2.070      & 163 & 20 & 24 & 2.0 & -0.60 & 19.9 & 0.302 & 0.291 & 0.008\\
\hline
\end{tabular}
\caption{%
We show the statistics of voids used to make the histogram in the figure \ref{mill_void}.
From left to right, we have redshifts, $z$, numbers of voids used to draw histograms, $N_{\rm v}$,
minimum numbers of void galaxies per void, $N_{\rm vg,min}$,
median numbers of void galaxies per void, $N_{\rm vg,med}$, 
linear galaxy biases, $b_L$, mean density contrasts of void, $\bar{\delta}_{\rm v}$,
mean Eulerian radii of voids, $\bar{R}_E\equiv(\overline{p_1p_2p_3})^{1/3}$ ($h^{-1}{\rm Mpc}$),
mean ellipticities of extracted voids, $\bar{\epsilon}^{\rm obs}$, and those of analytic PDF,
$\bar{\epsilon}^{\rm theory}$, and errors in the mean, $\Delta\epsilon^{\rm obs}$.
}%
\label{tb:mill_void}
\end{center}
\end{table*}

\subsection{Comments on the Void-Finder Dependence}
\label{sec:comments}

In this work, we have assumed that voids are the local minima of
density field with three dimensionally expanding volume.
In the Lagrangian picture, this is equivalent to 
having all the eigenvalues of the tidal tensor, $T_{ij}\equiv\partial_i\partial_j\Psi(\mathbf{q})$,
being negative (i.e., $\lambda_3\ge\lambda_2\ge\lambda_1<0$, and $\sum_{i=1}^3 \lambda_i=\delta_m$).

However, as \citet{lavaux/wandelt:2010} pointed out,
in a saddle-like density distribution, voids can contract along one of the axis,
while it expands along the rest of the axes.
\citet{lavaux/wandelt:2010} studied these spurious voids in the
Lagrangian picture together with their own void-finding algorithm (DIVA). 
They found that among those spurious voids, pancake voids
(i.e., two dimensionally expanding, one dimensionally contracting voids)
are as abundant as three dimensionally expanding genuine voids,
and the shape of the ellipticity distribution function of PL07 needs to be
modified.
Nevertheless, we focused our work on the three dimensionally expanding
voids, as the result of PL07 suggests that the void finding algorithm of
\citet{hoyle/vogeley:2002} mostly finds three dimensionally
expanding genuine voids. There seems to be an algorithm dependent
selection of the type of voids.

FN09 reported a slight preference for
prolate ellipsoids in the ellipticity PDF drawn from voids detected
in the SDSS DR5, while the analytic ellipticity PDF of PL07 shows a clear
preference for oblate ellipsoids.
FN09 used the best-fit ellipsoid of \citet{jang-condell/hernquist:2001}
to estimate the ellipticity of a void with a Monte-Carlo simulation.
For each realization, they calculated an inertia tensor for each void
from test particles randomly spread inside the void. In this way,
their measurements of ellipticity tend to be more sensitive to the
shape of the void boundary (i.e., $I_{ij}\propto p_ip_j$).
On the other hand, PL07 uses field galaxies {\it inside} the void
to calculate the inertia tensor.
An ellipticity calculated from void galaxies is more sensitive to
the shape of the tidal field around a density minimum, and therefore,
has a closer contact with the analytic ellipticity PDF of PL07.
Indeed, the mean effective volume of voids, $V_{\rm eff}=\frac{4\pi}3 p_1p_2p_3$,
extracted from the Millennium simulation with the method of PL07 is $\sim 4$ times smaller than that
of FN09 (i.e., most void galaxies are within $\sim 60\%$ of radius
from the void center).

\section{Discussion and Conclusion}
\label{sec:diss_con}

We have derived the ellipticity PDF of voids in the redshift space.
We have found that the redshift space distortion on the shape of voids
statistically increases the ellipticities of voids, and leaves
a prominent feature on the ellipticity PDF as a substantial
reduction in the probability of having voids with small ellipticity,
$\epsilon<-\frac{f\bar{\delta}}{3}$.
This characteristic cutoff in the ellipticity PDF can be used
as a probe of the growth rate, $f(a)\equiv\frac{d\ln D_+(a)}{d\ln a}$,
once the radial density profile of voids is better understood.
\citet{acquaviva/gawiser:2010} proposed a model independent test of GR
by checking the scale dependence/independence of the growth rate, $f(k,z)$
from a galaxy power spectrum at small scale, $P_g(k_S,z)$,
and large scale $P_g(k_L,z)$.
In principle, we can use the redshift space ellipticity PDF of a void
to measure the growth rate at different scales by binning
the observed voids into small and large sizes.

However, as we have shown in \S~\ref{sec:poisson_noise}, the biggest limiting factor for
the use of ellipticity PDF as a probe of cosmology lies in the Poisson noise
from a small number of $N_{\rm vg}$ inside a given void.
We have found that from a sample of galaxies with a minimum halo mass of
$2.6\times 10^{12}h^{-1}M_{\odot}$ for $z=0$ and $1$, and $8.6\times 10^{11}h^{-1}M_{\odot}$ for $2$,
we have $N_{\rm vg}\sim 20$ void galaxies per each identified void.
This small number of void galaxies per each void creates a significant
contamination of the resulting ellipticity PDF so that the shape of
the original PDF is almost washed-out.

As we have seen, the biggest limiting factor for
the use of ellipticity PDF as a probe of cosmology lies in the Poisson noise
from a small number of $N_{\rm vg}$ inside a given void.
We have found that from a sample of galaxies with a minimum halo mass of
$2.6\times 10^{12}h^{-1}M_{\odot}$ for $z=0$ and $1$, and $8.6\times 10^{11}h^{-1}M_{\odot}$ for $2$,
we have $N_{\rm vg}\sim 20$ void galaxies per each identified void.
This small number of void galaxies per each void creates a significant
contamination of the resulting ellipticity PDF so that the shape of
the original PDF is almost washed-out.

Nevertheless, there is a way to overcome the Poisson noise.
Recently, \citet{lavaux/wandelt:2011} proposed the stacking analysis of the void
ellipticity in the redshift space as a way to put a constraint on
the cosmological parameters via the Alcock Paczynski (AP) test
\citep{alcock/paczynski:1979}.
The AP test uses the deviation from the isotropy and homogeneity
of the observed structure, which is known to be spherical
(e.g., galaxy distributions at large scale), to determine $D_AH$.
For example, when one calibrates comoving distances
perpendicular, $r_{\perp}$, and parallel to the line of sight,
$r_{\parallel}$, from the angular, $\theta$, and redshift, $z$,
distribution of galaxies with
\begin{eqnarray}
  \theta&=&\frac{r_{\perp}(z)}{(1+z)D_A(z)},\\
  \delta z&=&\frac{r_{\parallel}(z)H(z)}{c},
\end{eqnarray}
one need to assume a reference cosmological model to calculate
$D^{\rm ref}_A(z)$ and $H^{\rm ref}(z)$.
If the observed structure has $r_{\perp}=r_{\parallel}$,
and we have observed values of the angular, $\theta$, and
the redshift, $z$, distributions of galaxies,
a combination of $D_A$ and $H$ can be constrained as follows,
\begin{eqnarray}
D_A(z)H(z)=\frac{c\delta z}{(1+z)\theta}
\end{eqnarray}
(see, e.g., \citet{shoji/jeong/komatsu:2009,kazin/sanchez/blanton:2012}
for the use of AP test on the two-dimensional power spectrum, $P(k,\mu)$).

In real space, the stacked void  with sufficient number of stacked void
galaxies should have zero ellipticity,
while the deviation from zero ellipticity in real space indicates the
deviation of the assumed cosmology from the true cosmology.
In redshift space, since the void is elongated toward the line of sight,
the stacked void has non-zero ellipticity, which can be a tell-tale of the
logarithmic growth rate, $f(a)=\frac{d\ln D_+(a)}{d\ln a}$.
Although some useful information of void ellipticity will be
lost by stacking, in this way, we can see the effect of redshift space
distortion as a source of anisotropy in the stacked
void ellipticity.
We think that the stacking analysis of the voids in redshift space
is potentially a powerful tool to probe the cosmology.
\acknowledgments
MS would like to thank Caroline Foster for letting us use their
void-finder, which was used in \S~\ref{sec:n-body} of this paper.
MS would also thank M. Kubota and E. Komatsu for their
warm hospitality and affectionate supervision.
The Millennium Simulation databases used in this paper and the 
web application providing on-line access to them were constructed as 
part of the activities of the German Astrophysical Virtual Observatory.
\appendix
\section{Calculation of $\kappa(\mathbf{x})$}
\label{sec:kappa}
A mapping between the real-space to the redshift space is
\begin{eqnarray}
x_1^s&=&x_1^r
\label{eq:x1_app}\\
x_2^s&=&x_2^r
\label{eq:x2_app}\\
x_3^s&=&x_3^r-f\mathbf{u}\cdot\hat{\mathbf{x}}_3^r,
\label{eq:x3_app}
\end{eqnarray}
where $f\equiv\frac{d\ln\delta(a)}{d\ln a}$ is the growth rate,
and we set the unit vector, $\hat{\mathbf{x}}^r_3$, to be along the
line of sight direction.
Here, $\mathbf{u}\equiv\frac{-\mathbf{v}}{\mathcal{H}f}$, where
$\mathbf{v}$ is the peculiar velocity of a void galaxy,
and $\mathcal{H}(z)$ is a comoving Hubble rate.

We define $\kappa(\mathbf{x}^r)$ to quantify the strength of
the redshift space distortion as follows
\begin{eqnarray}
\kappa(\mathbf{x}^r)\equiv
-\frac{f\hat{\mathbf{x}_3^r}\cdot\mathbf{u}(\mathbf{x}^r)}{x_3^r},
\end{eqnarray}
and we re-write eq. (\ref{eq:x3_app}) as
\begin{eqnarray}
x_3^s&=&x_3^r(1+\kappa(\mathbf{x}^r)).
\label{eq:x3b_app}
\end{eqnarray}

Here, we use a spherical coordinate such that
\begin{eqnarray}
x_1&=&x\sin\theta_x\cos\phi_x\\
x_2&=&x\sin\theta_x\sin\phi_x\\
x_3&=&x\cos\theta_x
\end{eqnarray}
and
\begin{eqnarray}
k_1&=&k\sin\theta_k\cos\phi_k\\
k_2&=&k\sin\theta_k\sin\phi_k\\
k_3&=&k\cos\theta_k,
\end{eqnarray}
where $x\equiv|\mathbf{x}|=\sqrt{x_1^2+x_2^2+x_3^2}$, $k\equiv|\mathbf{k}|=\sqrt{k_1^2+k_2^2+k_3^2}$, and
$\theta$ and $\phi$ are the azimuthal and inclination angles respectively.
\begin{eqnarray}
\kappa(\mathbf{x}^r)&\equiv&
-\frac{f\hat{\mathbf{x}_3^r}\cdot\mathbf{u}(\mathbf{x}^r)}{x_3^r}
=-\frac{f\hat{\mathbf{x}_3^r}\cdot\mathbf{\nabla}\Phi(\mathbf{x}^r)}{4\pi Ga^2\bar{\rho}x_3^r}
\nonumber\\
&=&-\frac{f}{4\pi Ga^2\bar{\rho}x_3^r}
\int\frac{d^3k}{(2\pi)^3}i\mathbf{k}\cdot\hat{\mathbf{x}}_3^r
\tilde{\Phi}(\mathbf{k})e^{i\mathbf{k}\cdot\mathbf{x}^r}
\nonumber\\
&=&\frac{f}{x_3^r}\int\frac{d^3k}{(2\pi)^3}i\mathbf{k}\cdot\hat{\mathbf{x}}_3^r
\frac{\tilde{\delta}(\mathbf{k})}{k^2}e^{i\mathbf{k}\cdot\mathbf{x}^r}
\nonumber\\
&=&\frac{f}{x_3^r}\int_0^{\infty}\frac{dk}{(2\pi)^3}ik\int_0^{2\pi}d\phi_k\int_{-1}^1
d\mu_k \mu_k\tilde{\delta}(\mathbf{k})e^{i\mathbf{k}\cdot\mathbf{x}^r},
\nonumber\\
\label{eq:kappa_app}
\end{eqnarray}
where $\mu_k\equiv\cos\theta_k$.
Here, $\mathbf{k}\cdot\mathbf{x}=kx\cos\gamma$, where
\begin{eqnarray}
\cos\gamma\equiv \cos\theta_k\cos\theta_x+\cos(\phi_k-\phi_x)\sin\theta_k\sin\theta_x.
\end{eqnarray}
\subsection{Spherical Void}

As for the spherical void of radius $R$ with homogeneous top-hat density contrast,
$\delta(\mathbf{x})$, we have
\begin{eqnarray}
\tilde{\delta}(\mathbf{k})=\bar{\delta}\int d^3x'e^{-i\mathbf{k}\cdot\mathbf{x}'}
=4\pi\bar{\delta}\int_0^Rdx~x^2e^{-ikx}=4\pi R^3\bar{\delta}\frac{j_1(kR)}{kR}.
\end{eqnarray}
Therefore, eq. (\ref{eq:kappa_app}) becomes
\begin{eqnarray}
\kappa(\mathbf{x}^r)=\frac{4\pi R^3f\bar{\delta}}{x_3^r}
\int_0^{\infty}\frac{dk}{(2\pi)^3}ik\frac{j_1(kR)}{kR}
\int_0^{2\pi}d\phi\int_{-1}^1d\mu_k \mu_k e^{i\mathbf{k}\cdot\mathbf{x}^r}.
\end{eqnarray}
We first integrate the angular portion of the integration,
$\int d\Omega_k \mu_k e^{i\mathbf{k}\cdot\mathbf{x}^r}$,
expanding the exponent into Legendre polynomials,
\begin{eqnarray}
e^{i\mathbf{k}\cdot\mathbf{x}^r}=\sum_l^{\infty}(-i)^l(2l+1)j_l(-kx)P_l(\cos\gamma),
\end{eqnarray}
and using the Spherical harmonics addition theorem (a.k.a., Legendre addition theorem),
\begin{eqnarray}
P_l(\cos\gamma)&=&
\frac{4\pi}{2l+1}\sum_{m=-l}^{l}(-1)^mY^m_l(\theta_k,\phi_k)Y^{-m}_l(\theta_x,\phi_x)
\nonumber\\
&=&\frac{4\pi}{2l+1}\sum_{m=-l}^{l}Y^m_l(\theta_k,\phi_k)\bar{Y}^m_l(\theta_x,\phi_x)
\nonumber\\
&=&P_l(\cos\theta_k)P_l(\cos\theta_x)
\nonumber\\
&+&2\sum_{m=1}^l\frac{(l-m)!}{(l+m)!}
P_l^m(\cos\theta_k)P_l^m(\cos\theta_x)\cos[m(\phi_k-\phi_x)],
\nonumber\\
\label{eq:addition_app}
\end{eqnarray}
where $P_l^m(x)$ is an associated Legendre polynomial, and it is
related to the unassociated Legendre polynomial such that
\begin{eqnarray}
P_l^m(x)&=&(-1)^m(1-x^2)^{m/2}\frac{d^m}{dx^m}P_l(x)
\nonumber\\
&=&\frac{(-1)^m}{2^ll!}(1-x^2)^{m/2}\frac{d^{l+m}}{dx^{l+m}}(x^2-1)^l.
\end{eqnarray}
Therefore,
\begin{eqnarray}
&&\int d\Omega_k \mu_k e^{i\mathbf{k}\cdot\mathbf{x}^r}
=\sum_l^{\infty}(-i)^l(2l+1)j_l(-kx)\int d\Omega_k \mu_kP_l(\cos\gamma)
\nonumber\\
&=&\sum_l^{\infty}\sum_{m=-l}^l(-i)^l(2l+1)j_l(-kx)\left(\frac{4\pi}{2l+1}\right)^{3/2}
\nonumber\\
&\times& Y^{-m}_l(\theta_x,\phi_x)\int d\Omega_k
Y^0_1(\theta_k,\phi_k)\bar{Y}^{-m}_l(\theta_k,\phi_k)
\nonumber\\
&=&\sum_l^{\infty}\sum_{m=-l}^l(-i)^l(2l+1)j_l(-kx)\left(\frac{4\pi}{2l+1}\right)^{3/2}
Y^{-m}_l(\theta_x,\phi_x)\delta_{m0}\delta_{l1}
\nonumber\\
&=&-3ij_1(-kx)\left(\frac{4\pi}{3}\right)^{3/2}Y^0_1(\theta_x,\phi_x)
=4\pi ij_1(kx)\cos\theta_x,
\label{eq:solid_angle_sphere_app}
\end{eqnarray}
where we used $Y^0_1(\theta,\phi)=\sqrt{\frac{3}{4\pi}}\cos\theta$ and
$j_l(-x)=(-1)^lj_l(x)$.

Finally, we obtain $\kappa$ for a spherical void as
\begin{eqnarray}
\kappa(\mathbf{x}^r)&=&\frac{4\pi R^3f\bar{\delta}}{x_3^r}
\int_0^{\infty}\frac{dk}{(2\pi)^3}ik\frac{j_1(kR)}{kR}
[4\pi ij_1(kx)\cos\theta_x]
\nonumber\\
&=&-\frac{2R^2f\bar{\delta}\cos\theta_x}{\pi x_3^r}
\int_0^{\infty}dk j_1(kR)j_1(kx)
\nonumber\\
&=&-\frac{2R^2f\bar{\delta}\cos\theta_x}{\pi x_3^r}\left(\frac{\pi x}{6R^2}\right)
=-\frac{f\bar{\delta}}{3}.
\label{eq:sphere_kappa_app}
\end{eqnarray}

\subsection{Spheroidal Void with azimuthal symmetry}
Here, we consider a spheroidal void with an azimuthal symmetry
(i.e., an ellipsoid with two of the three principal axes have the same
length, $a=b\ne c$).
We again choose a spherical coordinate, but with the ellipsoidal volume element
being projected on to spherical volume element such that
\begin{eqnarray}
x_1&=&a~r\sin\theta\cos\phi\\
x_2&=&a~r\sin\theta\sin\phi\\
x_3&=&c~r\cos\theta,
\end{eqnarray}
where $0\le r \le 1$ and $d^3x=a^2c~r^2dr~d\phi~d\cos\theta$.

First, let us derive the Fourier transform of the spheroidal void
with a homogeneous top-hat density contrast, $\delta(\mathbf{x})$ as follows,
\begin{eqnarray}
\tilde{\delta}(\mathbf{k})=\bar{\delta}\int d^3xe^{-i\mathbf{k}\cdot\mathbf{x}}
=a^2c\bar{\delta}\int_0^1r^2dr\int d\Omega e^{-i\mathbf{k}\cdot\mathbf{x}},
\end{eqnarray}
where
\begin{eqnarray}
\mathbf{k}\cdot\mathbf{x}=kr[a\sin\theta_k\sin\theta_x\cos(\phi_k-\phi_x)
+c\cos\theta_k\cos\theta_x].
\end{eqnarray}
Here, we define
\begin{eqnarray}
\cos\gamma_a&\equiv&\sin\theta_k\sin\theta_x\cos(\phi_k-\phi_x)\\
\cos\gamma_c&\equiv&\cos\theta_k\cos\theta_x,
\end{eqnarray}
and
\begin{eqnarray}
\cos\gamma&\equiv&\cos\gamma_a+\cos\gamma_c\\
\cos\eta&\equiv&-\cos\gamma_a+\cos\gamma_c,
\label{eq:cos_eta_app}
\end{eqnarray}
where $\cos\eta$ is obtained by changing a sign of $\theta_x$.
In terms of $\cos\gamma$ and $\cos\eta$, we have
\begin{eqnarray}
\mathbf{k}\cdot\mathbf{x}=\frac12 kr[(c+a)\cos\gamma+(c-a)\cos\eta].
\end{eqnarray}
Again, we use the spherical harmonics addition theorem (eq. (\ref{eq:addition_app})).
Note that due to the definition of eq. (\ref{eq:cos_eta_app}) and from
eq. (\ref{eq:addition_app}), we see that $P_l(\cos\eta)=P_l(\cos\gamma)$, and therefore,
\begin{eqnarray}
e^{-i\mathbf{k}\cdot\mathbf{x}}&=&\exp[-\frac12ikr(c+a)\cos\gamma]
\exp[-\frac12ikr(c-a)\cos\eta]\nonumber\\
&=&\left[\sum_{l=0}^{\infty}(-i)^l(2l+1)j_l\left(\frac12kr(c+a)\right)P_l(\cos\gamma)\right]
\nonumber\\
&\times&\left[\sum_{l'=0}^{\infty}(-i)^{l'}(2l'+1)j_{l'}\left(\frac12kr(c-a)\right)P_{l'}(\cos\gamma)\right]
\nonumber\\
&=&\sum_{l=0}^{\infty}\sum_{l'=0}^{\infty}(-i)^{l+l'}(2l+1)(2l'+1)
j_l\left(\frac12kr(c+a)\right)
\nonumber\\
&\times&j_{l'}\left(\frac12kr(c-a)\right)P_l(\cos\gamma)P_{l'}(\cos\gamma).
\nonumber\\
\end{eqnarray}
From the Legendre addition theorem, we have
\begin{eqnarray}
&&\int d\Omega_x P_l(\cos\gamma)P_{l'}(\cos\gamma)
\nonumber\\
&=&\frac{(4\pi)^2}{(2l+1)(2l'+1)}
\sum_{m=-l}^l\sum_{m'=-l'}^{l'}(-1)^mY^m_l(\theta_k,\phi_k)
Y^{m'}_{l'}(\theta_k,\phi_k)
\nonumber\\
&\times&\int_0^{2\pi}d\phi_x\int_{-1}^1d\cos\theta_x
Y^{-m}_l(\theta_x,\phi_x)\bar{Y}^{m'}_{l'}(\theta_x,\phi_x)
\nonumber\\
&=&\frac{(4\pi)^2}{(2l+1)(2l'+1)}
\sum_{m=-l}^l\sum_{m'=-l'}^{l'}(-1)^mY^m_l(\theta_k,\phi_k)
Y^{m'}_{l'}(\theta_k,\phi_k)\delta_{ll'}\delta_{-mm'},
\nonumber\\
\end{eqnarray}
then we have
\begin{eqnarray}
\int d\Omega_x~e^{-i\mathbf{k}\cdot\mathbf{x}}
&=&\sum_{l=0}^{\infty}\sum_{l'=0}^{\infty}(-i)^{l+l'}(2l+1)(2l'+1)
j_l\left(\frac12kr(c+a)\right)
\nonumber\\
&\times&j_{l'}\left(\frac12kr(c-a)\right)
\int d\Omega_x P_l(\cos\gamma)P_{l'}(\cos\gamma)
\nonumber\\
&=&16\pi^2\sum_{l=0}^{\infty}\sum_{m=-l}^l(-1)^{l+m}
j_l\left(\frac12kr(c+a)\right)j_l\left(\frac12kr(c-a)\right)
\nonumber\\
&\times&Y^m_l(\theta_k,\phi_k)Y^{-m}_l(\theta_k,\phi_k).
\end{eqnarray}

Therefore,
\begin{eqnarray}
\tilde{\delta}(\mathbf{k})
&=&a^2c\bar{\delta}\int_0^1r^2dr\int d\Omega e^{-i\mathbf{k}\cdot\mathbf{x}}
\nonumber\\
&=&16\pi^2 a^2c\bar{\delta}\sum_{l=0}^{\infty}\sum_{m=-l}^l(-1)^{l+m}
Y^m_l(\theta_k,\phi_k)Y^{-m}_l(\theta_k,\phi_k)
\nonumber\\
&\times&\int_0^1r^2drj_l\left(\frac12kr(c+a)\right)j_{l}\left(\frac12kr(c-a)\right),
\end{eqnarray}
and integrating over $r\in[0,1]$, we have
\begin{eqnarray}
\tilde{\delta}(\mathbf{k})&=&-\frac{8\pi^3a\bar{\delta}}{\sqrt{c^2-a^2}k^2}
\sum_{l=0}^{\infty}\sum_{m=-l}^l(-1)^{l+m}
Y^m_l(\theta_k,\phi_k)Y^{-m}_l(\theta_k,\phi_k)
\nonumber\\
&\times&\left[(c+a)J_{l-\frac12}\left(\frac12k(c+a)\right)
J_{l+\frac12}\left(\frac12k(c-a)\right)
\right.
\nonumber\\
&-&\left.(c-a)J_{l-\frac12}\left(\frac12k(c-a)\right)
J_{l+\frac12}\left(\frac12k(c+a)\right)
\right].
\end{eqnarray}
We Taylor expand the above equation for $\tilde{\delta}(\mathbf{k})$
about $\epsilon\simeq 0$ up to a linear order.
\begin{eqnarray}
\tilde{\delta}(\mathbf{k})&\simeq&
\frac{a\pi^3\bar{\delta}}{k^2}\sum_{l=0}^{\infty}
\frac{(ka)^{l+1/2}}{2^{2l-3/2}\Gamma(l+3/2)}\epsilon^l
\left[ka\epsilon J_{l+\frac12}(ka)+\{2+(l-1)\epsilon\}J_{l+\frac32}(ka)\right]
\nonumber\\
&\times&\sum_{m=-l}^l(-1)^{l+m}Y_l^m(\theta_k,\phi_k)Y_l^{-m}(\theta_k,\phi_k)
\end{eqnarray}
For a prolate ellipsoid, where $a<c$ and $\epsilon\equiv 1-\frac{a}{c}$,
\begin{eqnarray}
\tilde{\delta}(\mathbf{k})\simeq
4\pi\bar{\delta}a^3\left[
\frac{j_1(ka)}{ka}+\epsilon
\left(\frac{j_1(ka)}{ka}-j_2(ka)
\right)
\right]+\mathcal{O}(\epsilon^2).
\label{eq:delta1st_app}
\end{eqnarray}
For an oblate ellipsoid, where $a>c$ and $\epsilon\equiv 1-\frac{c}{a}$,
\begin{eqnarray}
\tilde{\delta}(\mathbf{k})\simeq
4\pi\bar{\delta}a^3\left[
\frac{j_1(ka)}{ka}-\epsilon
\left(\frac{j_1(ka)}{ka}-j_2(ka)
\right)
\right]+\mathcal{O}(\epsilon^2).
\label{eq:delta1st_app2}
\end{eqnarray}

Therefore, eq. (\ref{eq:kappa_app}) becomes
\begin{eqnarray}
\kappa(\mathbf{x})&=&\frac{4\pi a^3f\bar{\delta}}{x_3}
\int_0^{\infty}\frac{dk}{(2\pi)^3}ik
\left[\frac{j_1(ka)}{ka}\pm\epsilon
\left(\frac{j_1(ka)}{ka}-j_2(ka)\right)
\right]
\nonumber\\
&&\int d\Omega_k\cos\theta_k e^{i\mathbf{k}\cdot\mathbf{x}},
\label{eq:kappa_delta_app}
\end{eqnarray}
where
\begin{eqnarray}
\int d\Omega_k\cos\theta_k e^{i\mathbf{k}\cdot\mathbf{x}}
&=&\sum_{l=0}^{\infty}\sum_{l'=0}^{\infty}\sqrt{\frac{4\pi}{3}}
(-i)^{l+l'}(2l+1)(2l'+1)
j_l\left(-\frac12kr(c+a)\right)
\nonumber\\
&\times&j_{l'}\left(-\frac12kr(c-a)\right)
\int d\Omega_kY_1^0(\theta_k,\phi_k)P_l(\cos\gamma)P_{l'}(\cos\gamma).
\nonumber\\
\end{eqnarray}
Using the addition theorem given in eq. (\ref{eq:addition_app}),
we rewrite the above equation,
\begin{eqnarray}
\int d\Omega_k\cos\theta_k e^{i\mathbf{k}\cdot\mathbf{x}}
&=&\sum_{l=0}^{\infty}\sum_{l'=0}^{\infty}\sum_{m=-l}^{l}\sum_{m'=-l'}^{l'}
(4\pi)^2\sqrt{\frac{4\pi}{3}}
(-i)^{l+l'}
j_l\left(-\frac12kr(c+a)\right)
\nonumber\\
&\times&j_{l'}\left(-\frac12kr(c-a)\right)
(-1)^{m+m'}
Y_l^{-m}(\theta_x,\phi_x)Y_{l'}^{-m'}(\theta_x,\phi_x)
\nonumber\\
&\times&\int d\Omega_kY_1^0(\theta_k,\phi_k)
Y_l^m(\theta_k,\phi_k)Y_{l'}^{m'}(\theta_k,\phi_k).
\label{eq:solid_angle_app}
\nonumber\\
\end{eqnarray}
Here,
\begin{eqnarray}
&&j_l\left(-\frac12kr(c+a)\right)j_{l'}\left(-\frac12kr(c-a)\right)
\nonumber\\
&&\simeq2^{-2l'-\frac52}\pi\frac{(-1)^{l+l'}}{\Gamma(l'+\frac32)}(kra)^{l'-\frac12}
\epsilon^{l'}
\nonumber\\
&\times&\left[
\{2+(l+2l')\epsilon\}J_{l+\frac12}(kra)-kra\epsilon J_{l+\frac32}(kra)
\right]
\nonumber\\
&&=2^{-2(l'+1)}\frac{\sqrt{\pi}(-1)^{l+l'}}{\Gamma(l'+\frac32)}(kra)^{l'}
\epsilon^{l'}\left[
\{2+(l+2l')\epsilon\}j_l(kra)-kra\epsilon j_{l+1}(kra)
\right]
\nonumber\\
\label{eq:jjtaylor_app}
\end{eqnarray}
for a prolate ellipsoid ($c>a$), and
\begin{eqnarray}
&&j_l\left(-\frac12kr(c+a)\right)j_{l'}\left(-\frac12kr(c-a)\right)
\nonumber\\
&&\simeq2^{-2l'-\frac52}\pi\frac{(-1)^l}{\Gamma(l'+\frac32)}(kra)^{l'-\frac12}
\epsilon^{l'}\left[
(2-l\epsilon)J_{l+\frac12}(kra)+kra\epsilon J_{l+\frac32}(kra)
\right]
\nonumber\\
&&=2^{-2(l'+1)}\frac{\sqrt{\pi}(-1)^l}{\Gamma(l'+\frac32)}(kra)^{l'}
\epsilon^{l'}\left[
(2-l\epsilon)j_l(kra)+kra\epsilon j_{l+1}(kra)
\right]
\label{eq:jjtaylor_app2}
\end{eqnarray}
for an oblate ellipsoid ($c<a$).
In order to truncate the equation at linear
order in $\epsilon$, we have $l'\le 1$, such that
\begin{eqnarray}
&&j_l\left(-\frac12kr(c+a)\right)j_{l'}\left(-\frac12kr(c-a)\right)
\nonumber\\
&&\simeq\left\{
\begin{array}{l}
\frac12(-1)^l\left[
(2+l\epsilon)j_l(kra)-kra\epsilon j_{l+1}(kra)
\right]~~~~~~~~{\rm for}~l'=0\\
-\frac16(-1)^lkra\epsilon j_l(kra)
~~~~~~~~~~~~~~~~~~~~~~~~~~~~~~~~~~{\rm for}~l'=1\\
\mathcal{O}(\epsilon^2)~~~~~~~~~~~~~~~~~~~~~~~~~~~~~~~~~~~~~~~~~~~~~~~~~~~~{\rm otherwise}
\end{array}
\right.
\end{eqnarray}
for a prolate ellipsoid, and
\begin{eqnarray}
&&j_l\left(-\frac12kr(c+a)\right)j_{l'}\left(-\frac12kr(c-a)\right)
\nonumber\\
&&\simeq\left\{
\begin{array}{l}
\frac12(-1)^l\left[
(2-l\epsilon)j_l(kra)+kra\epsilon j_{l+1}(kra)
\right]~~~~~~~~{\rm for}~l'=0\\
\frac16(-1)^lkra\epsilon j_l(kra)
~~~~~~~~~~~~~~~~~~~~~~~~~~~~~~~~~~{\rm for}~l'=1\\
\mathcal{O}(\epsilon^2)~~~~~~~~~~~~~~~~~~~~~~~~~~~~~~~~~~~~~~~~~~~~~~~~~~~~{\rm otherwise}
\end{array}
\right.
\end{eqnarray}
for an oblate ellipsoid.

As for the integration over the solid angle, $\Omega_k$, we have
\begin{eqnarray}
&&\int d\Omega_kY_1^0(\theta_k,\phi_k)
Y_l^m(\theta_k,\phi_k)Y_{l'}^{m'}(\theta_k,\phi_k)
\nonumber\\
&&=\sqrt{\frac{3(2l+1)(2l'+1)}{4\pi}}
\left(
\begin{array}{ccc}
1 & l & l'\\
0 & 0 & 0
\end{array}
\right)
\left(
\begin{array}{ccc}
1 & l & l'\\
0 & m & m'
\end{array}
\right),
\end{eqnarray}
where the Wigner-3J symbol must satisfy the triangular inequality
such that $|1-l|\le l'\le 1+l$, and $0+m=-m'$ in order to have
a non-zero value.

As for eq. (\ref{eq:solid_angle_app}), since $l'\le 1$, we have a limited
number of non-zero terms.
We have $(l,l',m,m')=(0,1,0,0)$, $(1,0,0,0)$, $(2,1,-1,1)$, $(2,1,0,0)$ and
$(2,1,1,-1)$ as the non-zero terms.
We rewrite eq. (\ref{eq:solid_angle_app}) with a finite number of terms truncating
$\mathcal{O}(\epsilon^2)$ and higher as follows,
\begin{eqnarray}
\int d\Omega_k\cos\theta_k e^{i\mathbf{k}\cdot\mathbf{x}}
&\simeq&\frac{(4\pi)^2}{\sqrt{3}}i
\left[
\pm\frac16kra\epsilon j_0(kra)Y_0^0(\theta_x,\phi_x)Y_1^0(\theta_x,\phi_x)
\right.\nonumber\\
&+&\frac12\left\{
(2\pm\epsilon)j_1(kra)\mp kra\epsilon j_2(kra)
\right\}
Y_0^0(\theta_x,\phi_x)Y_1^0(\theta_x,\phi_x)
\nonumber\\
&\pm&\left.\frac16kra\epsilon j_2(kra)
\left\{
\sqrt{\frac35}Y_1^{-1}(\theta_x,\phi_x)Y_2^1(\theta_x,\phi_x)
\right.\right.
\nonumber\\
&+&\left.\left.\sqrt{\frac35}Y_1^1(\theta_x,\phi_x)Y_2^{-1}(\theta_x,\phi_x)
-\frac2{\sqrt{5}}Y_1^0(\theta_x,\phi_x)Y_2^0(\theta_x,\phi_x)
\right\}
\right]
\nonumber\\
\label{eq:solid_angle_spheroid1_app}
\end{eqnarray}
Here, in the limit of $\epsilon\to 0$, we recover
\begin{eqnarray}
\int d\Omega_k\cos\theta_k e^{i\mathbf{k}\cdot\mathbf{x}}
&=&\frac{(4\pi)^2}{\sqrt{3}}i j_1(kra)
Y_0^0(\theta_x,\phi_x)Y_1^0(\theta_x,\phi_x)
=4\pi i j_1(kra)\cos\theta_x,
\nonumber\\
\end{eqnarray}
as in eq. (\ref{eq:solid_angle_sphere_app}).

Finally, from eq. (\ref{eq:kappa_delta_app}) and (\ref{eq:solid_angle_spheroid1_app}),
and truncating $\mathcal{O}(\epsilon^2)$ and higher terms, we have
\begin{eqnarray}
\kappa(\mathbf{x})&=&\frac{4\pi a^3f\bar{\delta}}{x_3}
\int_0^{\infty}\frac{dk}{(2\pi)^3}ik
\left[\frac{j_1(ka)}{ka}
+\epsilon\left(\frac{j_1(ka)}{ka}-j_2(ka)\right)\right]
\nonumber\\
&\times&\int d\Omega_k\cos\theta_k e^{i\mathbf{k}\cdot\mathbf{x}}
\nonumber\\
&\simeq&
-\frac{8f\bar{\delta}a^3}{\sqrt{3}x_3}
\left[
\frac{r}6\epsilon Y_0^0(\theta_x,\phi_x)Y_1^0(\theta_x,\phi_x)
\int_0^{\infty}dk~kj_1(ka)j_0(kra)
\right.
\nonumber\\
&+&\frac{2+\epsilon}{2a}
Y_0^0(\theta_x,\phi_x)Y_1^0(\theta_x,\phi_x)
\int_0^{\infty}dk~j_1(ka)j_1(kra)
\nonumber\\
&-&\frac{r}2\epsilon
Y_0^0(\theta_x,\phi_x)Y_1^0(\theta_x,\phi_x)
\int_0^{\infty}dk~kj_1(ka)j_2(kra)
\nonumber\\
&+&\frac{r}6\epsilon
\left\{
\sqrt{\frac35}Y_1^{-1}(\theta_x,\phi_x)Y_2^1(\theta_x,\phi_x)
+\sqrt{\frac35}Y_1^1(\theta_x,\phi_x)Y_2^{-1}(\theta_x,\phi_x)
\right.
\nonumber\\
&-&\left.\frac2{\sqrt{5}}Y_1^0(\theta_x,\phi_x)Y_2^0(\theta_x,\phi_x)
\right\}
\int_0^{\infty}dk~kj_1(ka)j_2(kra)
\nonumber\\
&+&\frac{\epsilon}{a}Y_0^0(\theta_x,\phi_x)Y_1^0(\theta_x,\phi_x)
\int_0^{\infty}dk~j_1(ka)j_1(kra)
\nonumber\\
&-&\left.\epsilon Y_0^0(\theta_x,\phi_x)Y_1^0(\theta_x,\phi_x)
\int_0^{\infty}dk~kj_2(ka)j_1(kra)
\right].
\nonumber\\
\label{eq:kappa2_app}
\end{eqnarray}
Taking integrations over $k$ (see appendix \S\ref{sec:equations_app}),
\begin{eqnarray}
\kappa(\mathbf{x})&\simeq&-\frac{f\bar{\delta}}{3}\frac{ar\cos\theta_x}{x_3}
(1-\epsilon)
\nonumber\\
&=&-\frac{f\bar{\delta}}{3}\frac{a}{c}(1-\epsilon)
\nonumber\\
&=&-\frac{f\bar{\delta}}{3}(1-\epsilon)^2
\nonumber\\
&\simeq&-\frac{f\bar{\delta}}{3}(1-2\epsilon)
\label{eq:linear_kappa_app}
\end{eqnarray}
Similarly, for an oblate spheroid,
\begin{eqnarray}
\kappa(\mathbf{x})\simeq-\frac{f\bar{\delta}}{3}(1+2\epsilon)
\end{eqnarray}
As we compare against numerical calculations of $\kappa(\mathbf{x})$,
we see $\kappa(\mathbf{x})\simeq-\frac{f\bar{\delta}}{3}(1\mp\epsilon)$
fits the result better for both prolate and oblate spheroids.

We see how real space geometry of a given void affects the
peculiar velocity field within the void by comparing
Eqs.(\ref{eq:sphere_kappa_app}) and (\ref{eq:linear_kappa_app}).
As for a prolate spherical void, we set the longest principal
axis, $a<c\equiv a/(1-\epsilon)$, to lie along the line of sight.
As discussed by \citet{icke:1984}, originally aspherical voids
tend to become a sphere with a slower expansion speed along the
longest principal axis, and vice versa.
For a given prolate spheroidal void with its longest principal axis
along the line of sight, peculiar velocity along the largest axis become
smaller as we increase ellipticity.
Since the deformation of the void shape in the redshift space,
$\kappa(\mathbf{x})$, is proportional to the peculiar velocity,
an aspherical void in real space with the longest axis being along the line
of sight experiences less deformation compared to an originally
spherical void.

Also, we see that the degree of deformation of an aspherical void in the
redshift space, $\kappa(\mathbf{x})$, is independent of the
size of the void, but dependent only on the linear growth factor,
$f\equiv\frac{d\ln D}{d\ln a}$, the mean density contrast of the void,
$\bar{\delta}$, and the ellipticity of the void in the real space.
\section{List of $Y_l^m$, $j_l$ and Wigner-3J symbols used in appendix A}
\label{sec:equations_app}
Here, we collect the properties of the spherical harmonics,
$Y_l^m(\theta,\phi)$, spherical Bessel functions, $j_l(z)$,
Gamma functions, $\Gamma(x)$, and the Wigner-3J symbols,
$\left(
\begin{array}{ccc}
j_1 & j_2 & j_3\\
m_1 & m_2 & m_3
\end{array}
\right)$,
that are used in our calculations.

\begin{eqnarray}
Y_0^0(\theta,\phi)&=&\frac1{2\sqrt{\pi}}\\
Y_1^{-1}(\theta,\phi)&=&\frac12\sqrt{\frac3{2\pi}}\sin\theta e^{-i\phi}\\
Y_1^0(\theta,\phi)&=&\frac12\sqrt{\frac3{\pi}}\cos\theta\\
Y_1^1(\theta,\phi)&=&-\frac12\sqrt{\frac3{2\pi}}\sin\theta e^{i\phi}
\end{eqnarray}

\begin{eqnarray}
j_0(z)&=&\frac{\sin z}{z}\\
j_1(z)&=&\frac{\sin z}{z^2}-\frac{\cos z}{z}\\
j_2(z)&=&\left(\frac{3}{z^3}-\frac1{z}\right)\sin z-\frac3{z^2}\cos z\\
j_3(z)&=&\left(\frac{15}{z^4}-\frac6{z^2}\right)\sin z
-\left(\frac{15}{z^3}-\frac1{z}\right)\cos z
\end{eqnarray}

\begin{eqnarray}
\Gamma(1/2)&=&\sqrt{\pi}\\
\Gamma(3/2)&=&\frac{\sqrt{\pi}}{2}\\
\Gamma(5/2)&=&\frac{3\sqrt{\pi}}{4}\\
\Gamma(7/2)&=&\frac{15\sqrt{\pi}}{8}\\
\Gamma(9/2)&=&\frac{105\sqrt{\pi}}{16}
\end{eqnarray}

\begin{eqnarray}
\left(
\begin{array}{ccc}
1 & 0 & 1\\
0 & 0 & 0
\end{array}
\right)&=&-\frac{1}{\sqrt{3}}
\\
\left(
\begin{array}{ccc}
1 & 1 & 0\\
0 & 0 & 0
\end{array}
\right)&=&-\frac{1}{\sqrt{3}}
\\
\left(
\begin{array}{ccc}
1 & 2 & 1\\
0 & 0 & 0
\end{array}
\right)&=&\sqrt{\frac{2}{15}}
\\
\left(
\begin{array}{ccc}
1 & 2 & 1\\
0 & -1 & 1
\end{array}
\right)&=&-\frac{1}{\sqrt{10}}
\\
\left(
\begin{array}{ccc}
1 & 2 & 1\\
0 & 1 & -1
\end{array}
\right)&=&-\frac{1}{\sqrt{10}}
\end{eqnarray}

For $0<r<1$ and $a>0$,
\begin{eqnarray}
\int_0^{\infty}dk~j_1(ka)j_1(kra)&=&\frac{\pi r}{6a}\\
\int_0^{\infty}dk~k~j_2(ka)j_1(kra)&=&\frac{\pi r}{2a^2}\\
\int_0^{\infty}dk~k~j_1(ka)j_0(kra)&=&\frac{\pi}{2a^2}\\
\int_0^{\infty}dk~k~j_1(ka)j_2(kra)&=&0
\end{eqnarray}

\end{document}